\documentclass{appolb}
\usepackage{epsfig}
\usepackage{axodraw}
\begin{document}
\pagestyle{plain}

\newcount\eLiNe\eLiNe=\inputlineno\advance\eLiNe by -1

\title{\vspace{-20mm} 
\hspace{80mm} LU-TP 12-10\\\hspace*{78mm}
February 2012\vspace*{2cm}\\
Parton Cascades, Small $x$, and Saturation in High Energy Collisions
\footnote{Lecture notes combining two lectures and a contribution to
the celebration of Andrzej Bia\l as' birthday, presented at the LI Cracow School of
Theoretical Physics, Zakopane, June 2011}}
\author{G\"osta Gustafson\\
\address{Dept. of Theoretical Physics, Lund Univ., Sweden\\
S\"{o}lveg. 14A, 22362 Lund, Sweden\\
e-mail: gosta.gustafson@thep.lu.se}}
\maketitle

\begin{abstract}
These lecture notes are a combination of two lectures and a contribution to
the celebration of Andrzej Bia\l as' birthday at the LI Cracow School of
Theoretical Physics in June 2011. I here
discuss the dynamics of particle production in
high energy reactions. It includes parton cascades and hadronization in 
$e^+e^-$-ann., small $x$ evolution including the Double Leading Log 
approximation and the
BFKL equation, saturation at high densities and the BK equation, and 
finally the Lund Dipole Cascade 
model for high energy collisions, which is implemented in the DIPSY MC.
\end{abstract}

\tableofcontents

\section{Introduction}

In $e^+e^-$-annihilation the total hadronic cross
section is given by the probability to produce an initial $q\bar{q}$
pair, which is determined by QED. The effects of the strong interaction is here
only a correction with relative magnitude $\alpha_s/\pi$. The emission of a
gluon cascade is, however, essential for the properties of the
final state, although it does not change the total cross section. Here angular
ordering is crucial for the result, and the dipole formulation particularly 
convenient.

DIS and hadronic collisions are more complicated. 
In a high energy $ep$ or $pp$ collision the initial partons in a target proton
develop virtual parton cascades. A projectile can interact with any of the
partons in the cascade, which implies that the total cross section grows with
increasing collision energy. The development of the cascade in this ``initial 
state radiation'' determines the
inclusive total and elastic cross sections, but for exclusive final states
also ``final state radiation'' has to be added. In the initial state
radiation the virtualities are spacelike. The final state radiation
is more similar to the cascades in $e^+e^-$-ann.; it does not change
the inclusive cross sections, and the virtualities are timelike.
Thus in DIS and hadronic collisions we have two different problems, the total
cross section and the final state properties. There are also two different
hard scales, $Q^2$ and $s$, while in $e^+e^-$-ann. there is only one, $Q^2=s$.

At high energies and small $x$, gluon cascades and the $1/z$ pole in 
the splitting function
are most important. For large $Q^2$ this leads to the DLL approximation, and
for limited $Q^2$ to BFKL evolution.

The high density of partons in a proton also implies that 
at high energies the projectile may interact with more than one parton in the
target. As the total interaction probability must not be larger than one, the
effective gluon density must ``saturate''. At high energy the impact parameter
is related to the conserved angular momentum, $b\approx L/k$. The interaction
probability for fixed $b$ is therefore limited by 1, and a description of 
multiple interactions and saturation is most easy in impact
parameter space. 

The outline of these notes is first particle production in $e^+e^-$-ann. with
timelike cascades  and hadronization in Secs.~\ref{sec:timecasc} and 
\ref{sec:hadron}, followed by
small $x$ evolution in Sec.~\ref{sec:smallx} and saturation in 
Sec.~\ref{sec:sat}, and finally a discussion of dipole models for high energy
$pp$ collisions and DIS in Sec.~\ref{sec:dip}.

\section{Timelike cascades}
\label{sec:timecasc}

\subsection{Bremsstrahlung}

In classical electrodynamics the radiation of bremsstrahlung photons is given 
by the expression (see \eg Ref.~\cite{Jackson})
\begin{equation}
d n_\gamma \sim \frac{d^3 k}{\omega} \left|\int d^4 x\, \mathbf{j}(x) 
\mathbf{A}^*(x) \right|^2.
\label{eq:brems1}
\end{equation}
For a charged particle moving along the trajectory 
$\mathbf{x}= \mathbf{r}(t)$ we get the current (the charge is denoted $g$, as
the result is the same in $QCD$)
\begin{equation}
\mathbf{j}=g\, \mathbf{v}(t)\, \delta(\mathbf{x}- \mathbf{r}(t)).
\end{equation}
With a photon field $\mathbf{A} \sim \bar{\epsilon}
e^{-i(\omega t -\mathbf{k}\mathbf{x})}$ we find, after division and multiplication 
by $(1-\mathbf{n}\mathbf{v})$ and a partial integration, the amplitude
\begin{equation}
\mathcal{M} =\int d^4 x\, \mathbf{j}(x) \mathbf{A}^*(x)=
     i g \int dt\,\frac{dX}{dt}\,
e^{i\omega(t-\mathbf{n}\mathbf{r}(t))},
\label{eq:brems2}
\end{equation}
where
\begin{equation}
X=\frac{\bar{\epsilon}\mathbf{v}(t)}{\omega(1-\mathbf{n}\mathbf{v}(t))},
\,\,\,\,\,\,\mathbf{n}=\mathbf{k}/\omega.
\end{equation}
For soft emissions (small $\omega$) the exponential is approximately constant
in regions where $d\mathbf{v}/dt\ne 0$, which implies that
\begin{eqnarray}
\mathcal{M}&\propto& (X_f -X_i),\\
dn &\sim& \alpha\,\frac{d\omega}{\omega}\, d\Omega 
\left|\frac{\bar{\epsilon}\mathbf{v}_f}{1-\mathbf{n}\mathbf{v}_f}-
\frac{\bar{\epsilon}\mathbf{v}_i}{1-\mathbf{n}\mathbf{v}_i}\right|^2,
\label{eq:brems}
\end{eqnarray}
where $\mathbf{v}_i$ and $\mathbf{v}_f$ are the velocities before and after
the radiation. (For large $\omega$ the emission is, however, sensitive to 
details in the trajectory.)

\subsection{Dipole radiation}

For emission from pair production of a positive and a negative particle, moving 
along trajectories $\mathbf{r}_+$ and $\mathbf{r}_-$, we get the current
\begin{equation}
\left\{ \begin{array}{ll}
t<0:&\,\,\, \mathbf{j}=0 \\ 
t>0:&\,\,\, \mathbf{j}=+g\, \mathbf{v}_+(t)\, \delta(\mathbf{x}-
\mathbf{r}_+(t))-g \, \mathbf{v}_-(t)\, \delta(\mathbf{x}-
\mathbf{r}_-(t)).
\end{array}
\right.
\end{equation}
Thus we see that we get the same result as in Eq.~(\ref{eq:brems}), only 
with the replacements
$\mathbf{v}_f \rightarrow \mathbf{v}_+$ and $ \mathbf{v}_i \rightarrow
\mathbf{v}_-$.

In the cms system the matrix element for photon emission becomes
\begin{equation}
|\mathcal{M}|^2 \propto \frac{4}{\omega^2 \sin^2\theta}=\frac{(p_+ p_-)}{(p_+
  k)(p_- k)}
\label{eq:dipoleinv}
\end{equation}
We note that the last expression is relativistically invariant, and thus can
be used in any Lorentz frame. 

We can compare this result with the expressions from the relevant 
Feynman diagrams. The two factors in the denominator
in Eq.~(\ref{eq:dipoleinv}) correspond to the propagators 
$1/(p_+ +k)^2=1/[2(p_+ k)]$ and $1/(p_- +k)^2=1/[2(p_- k)]$ obtained
when the photon is emitted from the positive and negative parent
respectively. Coherent emission from the two parents give the ``dipole
formula'' in Eq.~(\ref{eq:dipoleinv}). Denoting the particles with
momenta $p_+$, $p_-$, and $k$
by the numbers 1, 2, and 3, and defining $s_{ij}=(p_i+p_j)^2$, we also get
(including a proper factor $1/\pi$)
\begin{equation}
dn= \frac{\alpha}{\pi} \frac{d\,s_{13}\, d\,s_{23}}{s_{13} \,s_{23}}=
\frac{\alpha}{\pi}\frac{d\,k_\perp^2}{k_\perp^2}\,d\,y
\label{eq:dipoleky}
\end{equation}
where
\begin{equation}
k_\perp^2=\frac{s_{13} \,s_{23}}{s}\,\,\,\mathrm{and}\,\,\,
y=\frac{1}{2}\ln\frac{s_{23}}{s_{13}}
\label{eq:kdef}
\end{equation}
represent the transverse momentum and the rapidity in the dipole rest frame.
For gluon emission in QCD we get the same expression with the finestructure 
constant $\alpha$ replaced by $N_c \alpha_s/2$. (For radiation from quarks
there is a suppression factor $(1-1/N_c^2)$, which is not present for dipoles
formed by gluon charges.)

\subsection{Angular ordering}

With the help of Eq.~(\ref{eq:dipoleinv}) the dipole emission can also be 
written
\begin{equation}
d\,n\sim \alpha\frac{d\,\omega}{\omega} d\,\Omega \frac{a_{12}}{a_{13}a_{23}}\,\,\,
\,\,\mathrm{with}\,\,\,a_{ij}\equiv 1-\mathbf{n}_i \mathbf{n}_j =1-\cos
\theta_{ij}, 
\end{equation}
where $\mathbf{n}_i$ is the direction of particle $i$, and $\theta_{ij}$ is
the angle between $\mathbf{n}_i$ and $\mathbf{n}_j$. As 
in Eq.~(\ref{eq:dipoleky}) particle 3 corresponds to the emitted photon or
gluon. The last factor can be rewritten in the form
\begin{equation}
\frac{a_{12}}{a_{13}a_{23}}=\frac{1}{2} \left[ \frac{a_{12}-a_{13}+a_{23}}{a_{13}a_{23}}
+(1 \leftrightarrow 2)\right] \equiv \frac{1}{2} \left[X_1+X_2\right]
\label{eq:angorder1}
\end{equation}
The first term in the parenthesis ($X_1$) is non-singular when
$a_{23}\rightarrow 0$. Averaging this term over the azimuth angle, $\phi$,  
around $\mathbf{n}_1$, keeping the polar angle $\theta_{13}$ fixed, we get
\begin{equation}
\frac{1}{2\pi} \int X_1 d\phi=\frac{2}{a_{13}} \theta(\theta_{12}-\theta_{13})
\label{eq:angorder2}
\end{equation}
A similar expression is obtained when averaging $X_2$ for fixed angle
$\theta_{23}$. Thus approximating $X_1$ and $X_2$ by these averages, the emission
corresponds to \emph{independent emission} from two emitters within the angular
ordered regions $\theta_{13} < \theta_{12}$ and $\theta_{23} < \theta_{12}$
respectively.  

\subsection{More gluons}

The emission of \emph{two gluons} is considerably more complicated. In a 
compressed form the lowest order result for $q\bar{q}gg$ final states takes 
three full pages in 
Ref.~\cite{Ellis:1980wv}. However, when the emissions are \emph{strongly 
ordered}, \emph{i.e.} when $p_4 \ll p_3 \ll W$, where $p_3$ and $p_4$ are the
gluon momenta, the result factorizes, and thus simplifies considerably. In a 
semiclassical picture the hardest gluon is emitted first from the $q\bar{q}$ 
dipole. This gluon carries away colour charge and thus changes the current
responsible for subsequent softer emissions. If the first emission produces 
\eg a red quark, a 
blue-antired gluon, and an antiblue antiquark, then the red-antired charges 
radiate coherently as a colour dipole formed by the quark and the gluon. In 
the rest frame of this dipole the distribution is also given by the expression 
in Eq.~(\ref{eq:dipoleky}). In the same way the blue and antiblue charges 
radiate coherently as a colour dipole formed by the gluon and the antiquark 
\cite{Azimov:1986sf}. (There is also a
colour-suppressed term corresponding to a dipole spanned between the quark and
the antiquark, with relative weight $-1/N_c^2$.) 
The emission of a gluon with transverse momentum $k_\perp$ is determined by
an average of the current in Eqs.~(\ref{eq:brems1}, \ref{eq:brems2}) over the
``Landau--Pomeranchuk
formation time'' $\tau \sim 1/k_\perp$. Thus the ordering of the
gluons is determined by their transverse momenta, when \emph{measured locally in
the emitting dipole rest frame}.

\begin{figure}
\includegraphics*[bb=75 608 510 770, width=12cm]{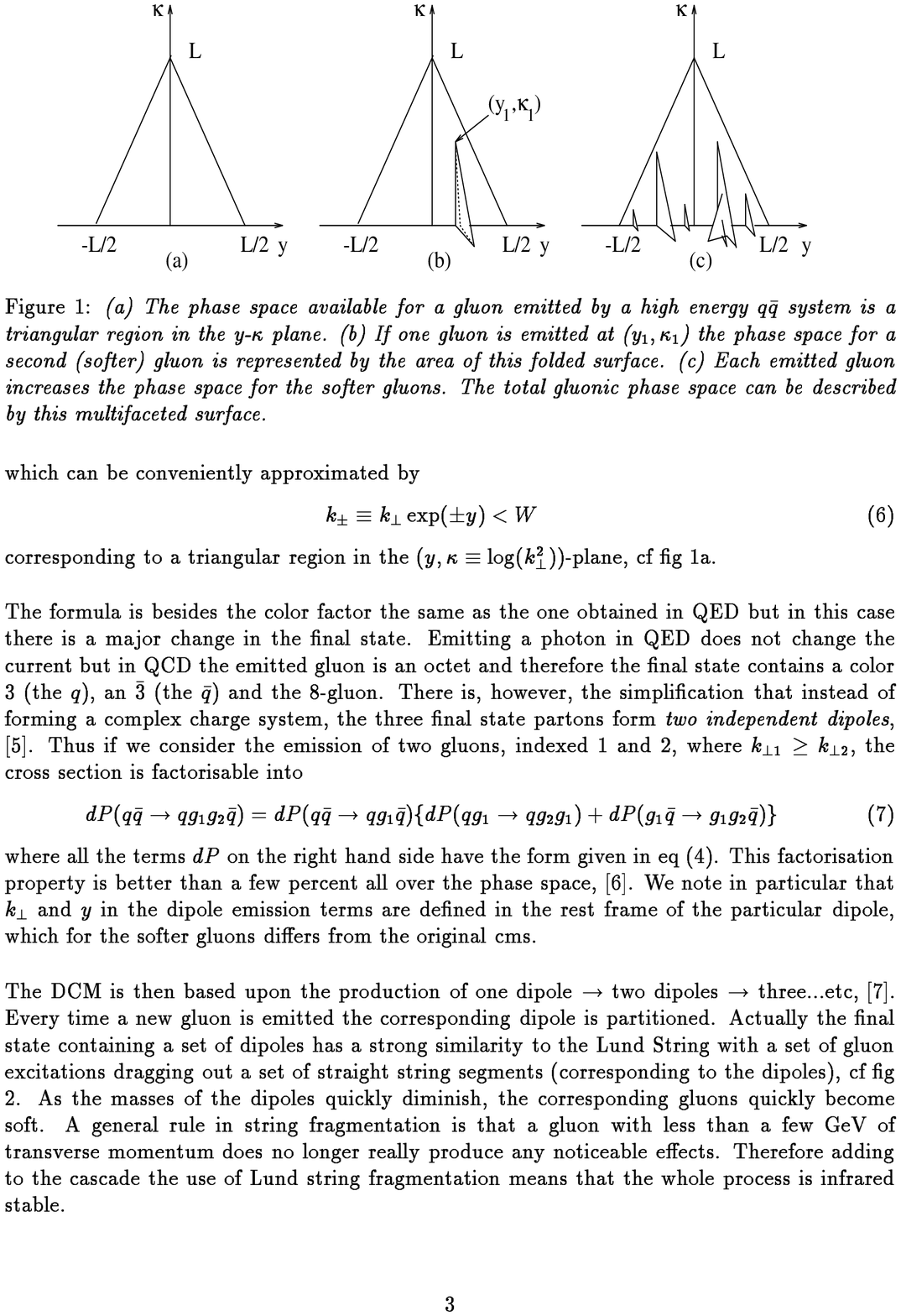}
\caption{\label{fig:fractal}
(a) The phase space for gluon emission in $e^+e^-$-ann. is a triangular 
region in the 
$(y,\kappa\equiv \ln k_\perp^2)$-plane. The height of the triangle 
is given by $L=\ln s$. (b) When one gluon is emitted at $(y_1, \kappa_1)$
the phase space for a second (softer) gluon is represented by the area of
this folded surface. (c) Each emitted gluon increases the phase space for
softer gluons. The total phase space is represented by this multifaceted
surface. }
\end{figure}

This result can be generalized so that the emission of \emph{many gluons} can
be described as a dipole cascade \cite{Gustafson:1987rq}. The phase space for
the emissions can be represented by the ($y,\kappa=\ln k_\perp^2$) diagram
shown in Fig.~\ref{fig:fractal}.
This formulation of the timelike parton cascade is implemented in the
\textsc{Ariadne} MC \cite{Lonnblad:1992tz}, which very successfully reproduces
experimental data from LEP and other $e^+e^-$ colliders.

The phase space in Fig.~\ref{fig:fractal} apparently has a fractal
structure. It is possible to define a fractal dimension given by 
$D=\sqrt{2N_c\alpha_s/\pi}$ \cite{Gustafson:1990qi}. As $\alpha_s(k_\perp^2)$ 
is running this is a so called multifractal. It has been discussed if this
feature is responsible for the ``intermittency'' signal observed in experimental
data. However, later it has been realized that a large fraction of the
observed effect is  related to BE correlations.

\section{Hadronization}
\label{sec:hadron}

\begin{figure}
 \begin{center}
  \includegraphics[angle=0,  scale=0.7]{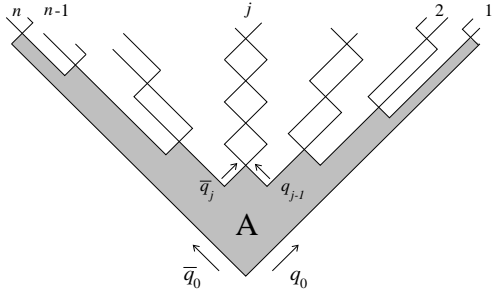}
  \caption{The hadronization of a high energy $(q_0,\bar{q}_0)$ system in a
    $(x,t)$ diagram.
  The hadrons can be ordered in ``rank'', 1, 2, \ldots\ $j$, \ldots\ $n$.
  This ordering agrees on average, but not in every case, with the ordering 
  in rapidity.} 
    \label{fig:solfjader}
 \end{center} 
\end{figure}

Quark confinement can be understood if the colour field is compressed to a
flux tube by a gluon condensate in vacuum. In an $e^+ e^-$-ann. event
a stringlike field is stretched out between a quark and an antiquark, and when
enough energy is stored in the field, it can break due by the production of a
new $q\bar{q}$ pair. A space-time picture of this process is shown in
Fig.~\ref{fig:solfjader}. 

In the Lund model the probability for
a definite final state is given by the product of a phase space
factor and the exponent of a constant times the space-time area spanned by the
string before it breaks, denoted by $A$ in Fig.~\ref{fig:solfjader} 
\cite{Andersson:1983jt}. This expression can be interpreted as a Wilson loop
integral, or an imaginary part of the string action. An
important feature of the result is boost invariance, which is also a
property of a homogenous longitudinal electric field.

\begin{figure}
\begin{center}
\includegraphics*[bb=150 460 500 670, width=9cm]{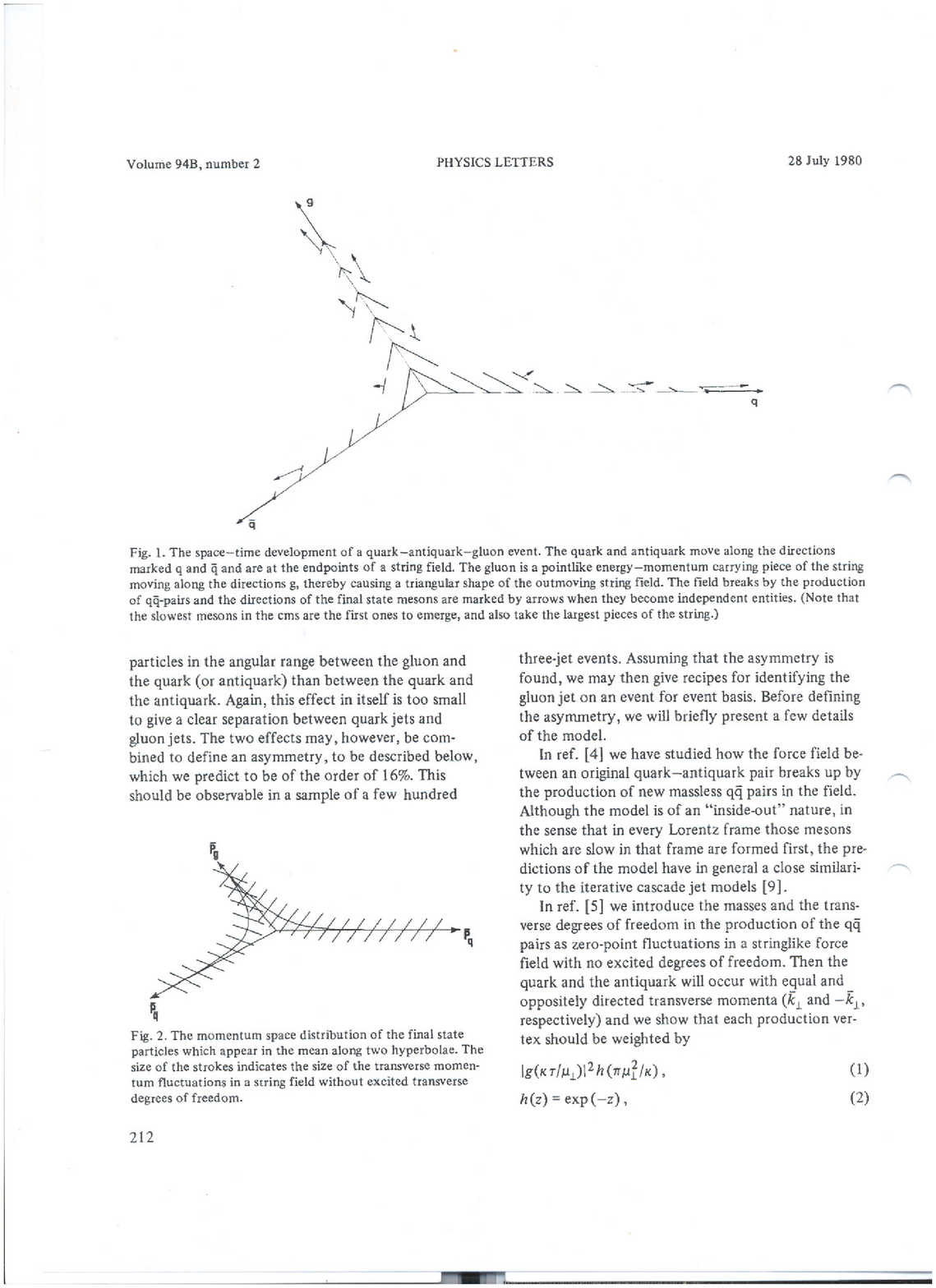}
  \caption{The space-time development of a quark-antiquark-gluon event.
The string is stretched from the quark to the antiquark via the gluon,
which moves like a pointlike kink carrying energy and momentum. The
string breaks by the production of new $q\bar{q}$ pairs, and the final state
contains three jets. Soft particles formed in between the jets get a boost
by the transverse motion of the string.}
    \label{fig:gluonjet}
 \end{center}
\end{figure}

A gluon carries colour and anticolour charges, and
in the Lund string hadronization model it behaves as a transverse excitation
on the stringlike field, stretched between a quark and an antiquark
\cite{Andersson:1979ij}. In a
three-jet event the string gets a transverse boost, and in the break-up the
hadrons are produced around two hyperbolae in momentum space, as shown in
Fig.~\ref{fig:gluonjet} \cite{Andersson:1980vk}. Thus fewer particles are 
produced in the angular
region opposite to the gluon jet, and this asymmetry was experimentally
confirmed, first by the JADE detector at the \textsc{Petra} collider
\cite{jade}. 

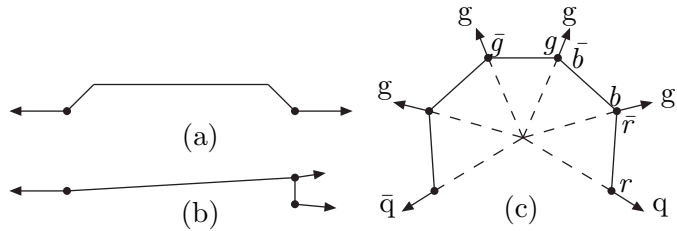
\begin{figure}
\begin{center}
\scalebox{1}{\mbox{
\begin{picture}(240,80)(-10,0)
\Line(15,10)(100,15)
\Line(100,15)(100,5)
\Line(15,40)(25,50)
\Line(25,50)(90,50)
\Line(90,50)(100,40)
\LongArrow(15,10)(-5,10)
\LongArrow(100,15)(110,16.5)
\LongArrow(100,5)(114,3.7)
\LongArrow(100,40)(120,40)
\LongArrow(15,40)(-5,40)
\Vertex(15,10){1.5}
\Vertex(100,15){1.5}
\Vertex(100,5){1.5}
\Vertex(15,40){1.5}
\Vertex(100,40){1.5}
\Text(65,1)[]{(b)}
\Text(65,30)[]{(a)}
\Line(152,10)(150,40)
\Line(150,40)(172,60)
\Line(172,60)(198,60)
\Line(198,60)(220,40)
\Line(220,40)(218,10)
\DashLine(185,30)(152,10){4}
\LongArrow(152,10)(141,3)
\DashLine(185,30)(150,40){4}
\LongArrow(150,40)(138,43)
\DashLine(185,30)(172,60){4}
\LongArrow(172,60)(168,70)
\DashLine(185,30)(198,60){4}
\LongArrow(198,60)(202,70)
\DashLine(185,30)(220,40){4}
\LongArrow(220,40)(232,43)
\DashLine(185,30)(218,10){4}
\LongArrow(218,10)(229,3)
\Vertex(152,10){1.5}
\Vertex(150,40){1.5}
\Vertex(172,60){1.5}
\Vertex(198,60){1.5}
\Vertex(220,40){1.5}
\Vertex(218,10){1.5}
\Text(185,3)[]{(c)}
\Text(237,4)[]{q}
\Text(224,12)[]{\textit{r}}
\Text(135,4)[]{\={q}}
\Text(240,45)[]{g}
\Text(225,35)[]{\textit{\={r}}}
\Text(220,46)[]{\textit{b}}
\Text(134,48)[]{g}
\Text(203,76)[]{g}
\Text(206,61)[]{\textit{\={b}}}
\Text(196,65)[]{\textit{g}}
\Text(164,76)[]{g}
\Text(176,65)[]{\textit{\={g}}}
\end{picture}
}}
\end{center}
\caption{\label{fig:soft} (a): A soft transverse gluon will soon lose
    its energy. The kink on the string is split in two corners and a straight
    string piece is stretched in a way similar to a one-dimensional string. 
    (b): Also for a collinear gluon the energy in the string between 
    the quark and the gluon is too small for a breakup of the string.
    (c): In a state with many gluons the string is stretched from
    the quark to the antiquark via the colour-ordered gluons, in the figure
    from red to antired, from blue to antiblue etc.}
\end{figure}

Gluon radiation is singular for soft and collinear emissions. A very
important feature of the string hadronization model is that it is 
\emph{infrared stable}. The motion of a soft transverse gluon is soon 
stopped by the tension in the attached strings. In the subsequent string 
motion the gluon kink is split into two corners, which do not
carry energy or momentum and which are connected by a straight string piece,
as shown in Fig.~\ref{fig:soft}a. The energy in
the small sections close to the quark and the antiquark is not sufficient for a
hadron, and all breakups will occur in the central string piece, which is
stretched and breaks up in the same way as the straight string in
Fig.~\ref{fig:solfjader}. 
The string motion with a collinear gluon is shown in Fig.~\ref{fig:soft}b,
and also here the effects of the gluon goes to zero in the collinear limit.

The situation in Fig.~~\ref{fig:gluonjet} can be directly generalized to 
many gluons. The string is here stretched from the quark to the antiquark
via the colour-ordered gluons, as shown in Fig.~\ref{fig:soft}c. 

\section{Spacelike cascades}
\label{sec:smallx}

\begin{figure}
\begin{center}
\scalebox{.92}{\mbox{
\begin{picture}(140,170)(-20,5)
  \Text(-10,15)[]{\large $P_a$}
  \Line(0,15)(40,15)
  \Line(40,20)(70,20)
  \Line(40,15)(70,15)
  \Line(40,10)(70,10)
  \GOval(40,15)(10,7)(0){1}
  \Line(50,20)(60,40)\Text(47,34)[]{\large $k_{0}$}
    \Line(60,40)(90,40)\Text(100,40)[]{\large $q_{1}$}
  \Line(60,40)(70,70)\Text(55,55)[]{\large $k_{1}$}
    \Line(70,70)(105,70)\Text(115,70)[]{\large $q_{2}$}
  \Line(70,70)(75,100)\Text(65,85)[]{\large $k_{2}$}
    \Line(75,100)(110,100)\Text(120,100)[]{\large $q_{3}$}
  \Line(75,100)(75,130)\Text(68,115)[]{\large $k_{3}$}
    \Line(75,130)(105,130)\Text(120,130)[]{\large $q_{4}$}
  \Photon(45,160)(75,130){4}{3}
  \Text(35,165)[]{\large $Q^2$}
  \DashLine(95,130)(105,120){2}
  \DashLine(75,115)(85,115){2}
  \DashLine(85,70)(95,80){2}
  \DashLine(75,40)(85,50){2}
  \DashLine(65,55)(75,55){2}

\end{picture}}}
\end{center}
\caption{A DIS event with ISR, solid lines, and FSR, dotted lines. Virtual
  links are denoted $k_i$ and real emissions $q_i$.}
\label{fig:ISR}
\end{figure}
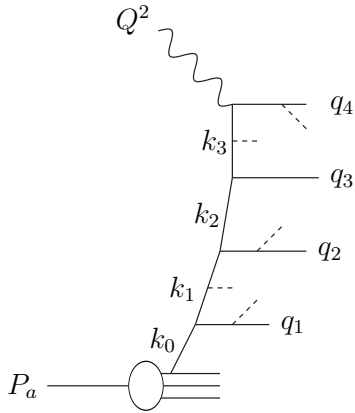

As discussed in the introduction, DIS and hadronic 
collisions are more complicated than $e^+e^-$-ann.. There are two
separate scales, $Q^2$ and $s$, and two separate problems: inclusive cross
sections and final state properties. The ladder leading up to the hard
interaction (solid lines in Fig.~\ref{fig:ISR}) represents the increased 
parton density in the initial state, and 
determines the inclusive total and elastic cross sections.
The parton links, $k_i$, in these cascades have spacelike momenta, and only those 
branches in the cascade, which interact with the projectile, can come
on shell and produce real final state particles. For exclusive final states 
also final state radiation has to be added (dashed lines in 
Fig.~\ref{fig:ISR}). This phase
is more similar to the cascades in $e^+e^-$-ann., with timelike virtualities
and conservation of probability.

For high $Q^2$ and not too small $x$, the ladder is described by ordered DGLAP
evolution, where $k_{\perp i} > k_{\perp i-1}$ and the vertices are determined by
the quark and gluon splitting functions. For small $x$,
gluon ladders are most important, and the  evolution dominated by
the $1/z$ pole in the gluon splitting function. This pole represents soft 
emissions, where each step in the ladder corresponds to a large step in
rapidity. In this section I will discuss small $x$ evolution in a
semiclassical framework, based on Weizs\"acker-Williams method of virtual 
quanta.  At high energies more than one
parton in the projectile or the target may interact. This problem will be
discussed in Sec.~\ref{sec:sat}.

\subsection{Weizs\"acker-Williams method of virtual quanta}

\begin{figure}
\begin{center}
\scalebox{0.8}{\mbox{
\begin{picture}(360,140)(0,0)
\BCirc(70,70){38}
\Vertex(70,70){2}
\LongArrow(40,70)(20,70)
\LongArrow(100,70)(120,70)
\LongArrow(70,100)(70,120)
\LongArrow(70,40)(70,20)
\Text(74,120)[l]{$\mathbf{E}$}
\Text(60,70)[]{+}

\Oval(190,70)(38,5)(0)
\Vertex(190,70){2}
\LongArrow(190,94)(190,126)
\LongArrow(190,46)(190,14)
\LongArrow(190,70)(216,70)
\Text(194,120)[l]{$\mathbf{E}_\perp$}
\Text(221,70)[l]{$\mathbf{v}$}

\DashLine(250,70)(335,70){5}
\CArc(322,65)(25,90,135)
\CArc(322,115)(25,225,270)
\CArc(315,90)(9,150,205)

\LongArrow(330,80)(330,90)
\LongArrow(330,80)(330,70)
\Text(334,80)[l]{$\mathbf{r}$}

\Text(70,10)[]{\emph{rest frame}}
\Text(240,10)[]{\emph{boosted frame}}

\end{picture}
}}
\end{center}
\caption{A Coulomb field in a boosted frame is compressed to a flat pancake}
\label{fig:WW}
\end{figure}
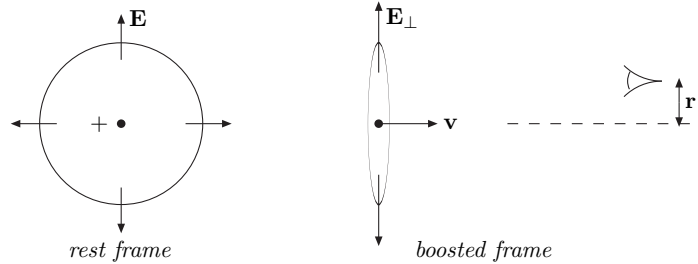

A Coulomb field, which is boosted to high velocity, is contracted to a flat 
pancake with a 
dominantly transverse electric field. The pulse will be very short in time, 
and can be approximated by a $\delta$-function:
\begin{equation}
\mathbf{E}_\perp \sim g\frac{\mathbf{r}}{r^2}\, \delta(t),
\label{eq:Efield}
\end{equation}
Here $\mathbf{r}$ is the 
(two-dimensional) distance between the position of the central charge and
the point of observation (see Fig. \ref{fig:WW}).
The frequency distribution is given by the Fourier transform,
and consequently approximately constant as a function of $\omega$, 
$E_\perp(\omega) \sim g/r$.

The electric field is also associated with an orthogonal transverse
magnetic field, with the same magnitude. The energy density in the pulse is
therefore given by
\begin{equation}
I(\omega) = E_\perp B_\perp \approx E_\perp^2(\omega)\sim g^2 \,\frac{1}{r^2}.
\label{eq:WW1}
\end{equation}
The density of photons, or gluons, seen by an observer at point $\mathbf{r}$, 
is obtained by dividing by the energy of a photon, and thus given by
\begin{equation}
dn \sim g^2\,\frac{d^2r}{r^2}\,\frac{d\omega}{\omega} \sim 
g^2\,\frac{d^2q_\perp}{q_\perp^2}\,\frac{d\omega}{\omega}.
\label{eq:WW2}
\end{equation}
In the last expression we used that the (twodimensional) Fourier transform of
a wavefunction proportional to $1/r$ is given by $1/q_\perp$.

\subsection{Dipoles in spacelike cascades}
\label{sec:dipcasc}

\begin{figure}
\scalebox{0.8}{\mbox{
\begin{picture}(125,72.5)(-10,-20)
\Line(5,5)(100,5)
\Line(5,5)(38,45.5)
\Line(38,45.5)(100,5)
\Vertex(5,5){2}
\Vertex(100,5){2}
\Vertex(38,45.5){2}
\LongArrow(38,45.5)(60,67.5)
\LongArrow(38,45.5)(56.3,34)
\LongArrow(38,45.5)(78.3,56)
\Text(80,59)[lt]{$\mathbf{E}$}
\Text(35.5,45)[rb]{$\mathbf{z}$}
\Text(50,2.5)[t]{$R$}
\Text(21.5,25.5)[rb]{$r_1$}
\Text(76.5,25)[b]{$r_2$}
\Text(49,65.5)[b]{$\mathbf{E}_1$}
\Text(56,40)[l]{$\mathbf{E}_2$}
\Text(1,10)[r]{$red$}
\Text(104,10)[l]{$\overline{red}$}
\Text(5,2.5)[t]{$\mathbf{x}$}
\Text(100,2.5)[t]{$\mathbf{y}$}
\Text(65,-18)[]{{\large (a)}}
\end{picture}
}}
\hspace{8mm}
\scalebox{0.65}{\mbox{
\begin{picture}(340,80)(0,-30)
\Vertex(10,80){2}
\Vertex(10,0){2}
\Text(5,80)[]{$\mathrm{x}$}
\Text(5,0)[]{$\mathrm{y}$}
\Line(10,80)(10,0)
\LongArrow(30,40)(60,40)
\Vertex(100,80){2}
\Vertex(100,0){2}
\Vertex(120,50){2}
\Text(95,80)[]{$\mathrm{x}$}
\Text(95,0)[]{$\mathrm{y}$}
\Text(128,50)[]{$\mathrm{z}$}
\Line(100,80)(120,50)
\Line(120,50)(100,0)
\LongArrow(140,40)(170,40)
\Vertex(205,80){2}
\Vertex(225,50){2}
\Vertex(233,30){2}
\Vertex(205,0){2}
\Text(200,80)[]{$\mathrm{x}$}
\Text(200,0)[]{$\mathrm{y}$}
\Text(233,50)[]{$\mathrm{z}$}
\Text(241,30)[]{$\mathrm{w}$}
\Line(205,80)(225,50)
\Line(225,50)(233,30)
\Line(233,30)(205,0)
\LongArrow(255,40)(285,40)
\Line(310,80)(320,70)
\Line(320,70)(302,63)
\Line(302,63)(325,58)
\Line(325,58)(330,50)
\Line(330,50)(320,43)
\Line(320,43)(338,30)
\Line(338,30)(310,35)
\Line(310,35)(310,0)
\Vertex(310,80){2}
\Vertex(320,70){2}
\Vertex(302,63){2}
\Vertex(325,58){2}
\Vertex(330,50){2}
\Vertex(320,43){2}
\Vertex(338,30){2}
\Vertex(310,35){2}
\Vertex(310,0){2}
\Text(170,-29)[]{{\large (b)}}
\end{picture}
}}
\caption{(a) The transverse colour-electric field in a colour dipole.
(b) Gluon emission splits the dipole into two dipoles. Repeated
emissions give a cascade, which produces a chain of dipoles.}
\label{fig:dipolesplit}
\end{figure}
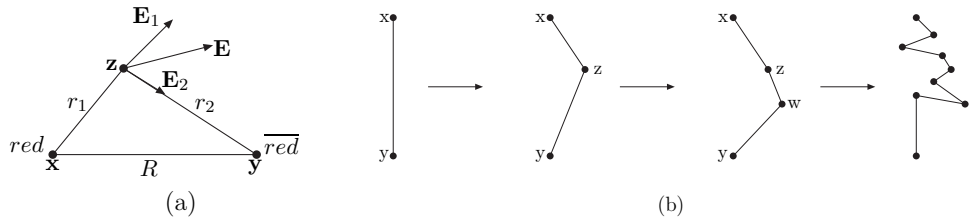

A proton is colour neutral, and the colour field from a parton is always 
screened by a corresponding anticharge. Let us study the field from a colour
dipole formed by a charge at $\mathbf{x}$ and an anticharge at $\mathbf{y}$ in
the transverse plane. The transverse field from these charges in point
$\mathbf{z}$ is given by (\emph{cf} Eq.~(\ref{eq:Efield}) and 
Fig.~\ref{fig:dipolesplit}(a)
\begin{equation}
\mathbf{E} = \mathbf{E}_1 + \mathbf{E}_2 \propto\frac{\mathbf{r}_1}{r_1^2}-
\frac{\mathbf{r}_2}{r_2^2},
\label{eq:dipolefield}
\end{equation}
where $\mathbf{r}_1=\mathbf{z}-\mathbf{x}$ and
$\mathbf{r}_2=\mathbf{z}-\mathbf{y}$. 
Defining $Y=\ln \omega$ and $\mathbf{R}=\mathbf{r}_1-\mathbf{r}_2$, we find 
in analogy with Eq.~(\ref{eq:WW2}) the gluon density in point $\mathbf{z}$:
\begin{equation}
\frac{d\,n}{dY\, d^2z} \propto \mathbf{E}^2 \propto 
\left(\frac{\mathbf{r}_1}{r_1^2}-\frac{\mathbf{r}_2}{r_2^2}\right)^2=
\frac{R^2}{r_1^2\cdot r_2^2}
\label{eq:transversedipole}
\end{equation}
We note that for small $r_1$ we have $R^2\approx r_2^2$, and 
Eq.~(\ref{eq:transversedipole}) corresponds to a pure Coulomb field $\propto
1/r_1^2$ from the charge in $\mathbf{x}$, while at larger distances, where 
$R\ll r_1 \approx r_2$, the field is screened and falls off $\sim 1/r^4$.

The essential difference between QCD and QED is that the emitted gluon carries
away colour charge. Thus, if \eg the gluon with colour $r\bar{b}$ is emitted
from an originally $r\bar{r}$ dipole, the originally red charge is changed to
blue, and the dipole is changed to a system of two dipoles, a $b\bar{b}$ dipole
formed by the originally red (but now blue) charge and the antiblue charge in
the emitted gluon, and a
$r\bar{r}$ dipole between this gluon and the original anticharge. In the large
$N_c$ limit these dipoles can emit softer gluons independently. The number of
dipoles increase as a \emph{cascade} to smaller and smaller rapidities $Y$, as
indicated in Fig.~\ref{fig:dipolesplit}b.

We note that the density proportional to $d\omega/\omega = dY$ corresponds to 
the $1/z$ pole in the $q\rightarrow qg$ and $g\rightarrow gg$ splitting
functions, which dominate the parton distribution for very small $x$.

\subsection{Double Leading Log approximation}

As mentioned the gluon emission is suppressed when $r_1$ and $r_2$ are larger
than $R$, and
the distribution in Eq.~(\ref{eq:transversedipole}) can be separated in a way
very similar to the angular ordering in the timelike cascade. We split the
expression in Eq.~(\ref{eq:transversedipole}) in the same way as in
Eq.~(\ref{eq:angorder1}):
\begin{equation}
\frac{R^2}{r_1^2\cdot r_2^2}=\frac{1}{2} \left[\frac{R^2-r_1^2+r_2^2}{r_1^2\,r_2^2}
+(1 \leftrightarrow 2)\right] \equiv \frac{1}{2} \left[X_1+X_2\right].
\label{eq:diporder1}
\end{equation}
Here the first term in the parenthesis ($X_1$) is non-singular when
$r_2\rightarrow 0$ (and $r_1^2\rightarrow R^2$). Averaging this term over 
the azimuth angle, $\phi$, around $\mathbf{x}$, keeping $r_1$ fixed,  we get
\begin{equation}
\frac{1}{2\pi} \int X_1 d\phi=\frac{2}{r_1^2} \theta(R-r_1).
\label{eq:diporder2}
\end{equation}
Thus the gluon emission in Eq.~(\ref{eq:transversedipole}) can be approximated
by
\begin{equation}
\frac{d\,n}{d^2r \,dY} \approx \frac{\bar{\alpha}}{2\pi}
\left[\frac{d^2 r_1}{r_1^2} \theta(R-r_1) +\frac{d^2 r_2}{r_2^2}
  \theta(R-r_2)\right],
\label{eq:diporder3}
\end{equation}
where we have included the proper numerical factor, and used the notation
$\bar{\alpha}=N_c \alpha_s/\pi$.
This corresponds to the independent emission from two single charges, confined
within the regions $r_i<R$. Thus the dipoles are ordered in size;
the daughter dipole is smaller than her parent. 

A probe with resolution $Q^2$ can ``see'' dipoles in a target with size
$r>1/Q$, while smaller dipoles are not resolved. Although non-ordered dipoles
are not totally excluded in the exact expression in 
Eq.~(\ref{eq:transversedipole}),
the approximation in Eq.~(\ref{eq:diporder3}) implies that the dipoles in a 
typical cascade become smaller and smaller. Therefore, for \emph{large $Q^2$} 
ordered emission chains dominate, in which $1/Q <\ldots< r_i<r_{i-1}< \ldots <
R$, where $R$ is the size of the initial dipole in the cascade.
Calculating the density of dipoles with size $r$ at rapidity $y$ in such a
cascade, we get first 
a contribution from emissions directly from the original dipole:
\begin{equation}
direct\,\, contribution:\,\,\,\bar{\alpha}\frac{d\,r^2}{r^2} dy \,\,\,\,\,\,\,
\,\,\,\,\,\,\,\,\,\,\,\,\,\,\,\,\,\,\,\,\,\,\,\,\,\,\,\,\,\,\,\,\,\,\,\,\,\,\,
\,\,\,\,\,\,\,\,\,\,\,\,\,\,\,\,\,\,\,\,\,\,\,\,\,\,\,\,\,\,\,\,\,\,\,\,\,\,\,
\,\,\,\,\,
\end{equation}
A two-step contribution is obtained by first emitting a dipole with size $r_1$
at rapidity $y_1$, which then emits the observed dipole at a lower rapidity
$y<y_1$:  
\begin{equation}
2\,\, steps:\,\,\bar{\alpha}\frac{d\,r^2}{r^2} dy
\int_r^R\bar{\alpha}\frac{d\,r_1^2}{r_1^2} \int_y^Y dy_1 =
\bar{\alpha}\frac{d\,r^2}{r^2} dy[\bar{\alpha} \ln(R^2/r^2) (Y-y)]
\end{equation}
Calling the square parenthesis $X$, we get in three steps
\begin{equation}
3\,\, steps:\,\,\bar{\alpha}\frac{d\,r^2}{r^2} dy
\int_r^R\bar{\alpha}\frac{d\,r_1^2}{r_1^2} \int_y^Y dy_1 
\int_{r}^{r_1}\bar{\alpha}\frac{d\,r_2^2}{r_2^2} \int_{y}^{y_1} dy_2 
=\bar{\alpha}\frac{d\,r^2}{r^2} dy\cdot\frac{1}{2^2} X^2.
\end{equation}
Summing contributions from $n$ steps, 
with $n=1 \ldots \infty$, gives the initial distribution from a single step
with the extra factor
\begin{equation}
\sum_n \frac{1}{(n!)^2} X^n =I_0(2\sqrt X).
\label{eq:bessel}
\end{equation}
Here $I_0$ is a Bessel function, which for large arguments grows like an
exponential. With $Y-y= \ln (1/x)$ we therefore get the result
\begin{equation}
\frac{d\,n}{d^2r \,dY} \sim\exp(2\sqrt{\bar{\alpha} \ln(R^2/r^2) \ln (1/x)}).
\end{equation} 
This result represents the
\emph{Double Leading Log}, or DLL, approximation, valid for small $r$ (meaning 
large $Q^2$) and small $x$, when both logarithms are large. 
(We have here
assumed a constant coupling $\bar{\alpha}$. For a running coupling $\propto 
1/\ln (1/\Lambda^2 r^2)$, $\ln (R^2/r^2)$ is replaced by 
$\ln [\ln (1/\Lambda^2 r^2)/\ln (1/\Lambda^2 R^2)]$.)

\subsection{BFKL evolution}
\label{sec:bfkl}

When $x$ is small but $Q^2$ not large, the integral over the ordered
dipoles, $r_i$, which leads to the factor $(\ln (R^2/r^2))^n/n!$ in
Eq.~(\ref{eq:bessel}), becomes small for large $n$. Although suppressed, 
unordered dipole chains become important,
in which some $r_i$ may be larger than $r_{i-1}$. We must then use the
full expression for
dipole emission in Eqs.~(\ref{eq:transversedipole}, \ref{eq:diporder1}),
instead of the approximation in Eq.~(\ref{eq:diporder3}). 

If the rapidity interval $Y-y=\ln(1/x)$ is increased by an amount $\delta Y$,
the density of dipoles $\mathcal{F}(Y, r^2)$ will change in the following
way: $\mathcal{F}$ can increase if a dipole with size $r'$ splits forming a
dipole $r$ within the interval $\delta Y$ (a gain term), and it can decrease 
if a dipole of size $r$ splits into two new dipoles (a loss term). This gives 
the following differential equation:
\begin{equation}
\frac{\partial \mathcal{F}(Y, r^2)}{\partial Y} =
\frac{\bar{\alpha}}{2\pi} \left\{ 
\int \frac{d^2 r'\cdot r'^2}{r^2(\mathbf{r}- \mathbf{r}')^2}
\mathcal{F}(Y,r'^2)\cdot 2 -
\int \frac{d^2 r'\cdot r^2}{r'^2(\mathbf{r}- \mathbf{r}')^2}
\mathcal{F}(Y,r^2)\right\}.
\label{eq:bfkl1}
\end{equation}
(The gain term has a factor 2, because 
when a dipole splits one or the other daughter can have size $r$.)

This equation is equivalent to the LL BFKL equation, conventionally formulated 
in transverse momentum space \cite{Mueller:1993rr}.
An important feature is here that the singularity in the gain term at 
$\mathbf{r}'= \mathbf{r}$ is compensated by the singularities in the loss term
at $\mathbf{r}'=0$ and $\mathbf{r}'= \mathbf{r}$. To see this more clearly, 
we note that the integrand in the loss term is symmetric under the exchange
$\mathbf{r}'\rightarrow \mathbf{r}- \mathbf{r}'$. We can therefore make
the following replacement:
\begin{equation}
\frac{1}{r'^2(\mathbf{r}- \mathbf{r}')^2} = 
\left[\frac{1}{r'^2}+\frac{1}{(\mathbf{r}-\mathbf{r}')^2}\right] 
    \frac{1}{r'^2+(\mathbf{r}- \mathbf{r}')^2 }
    \rightarrow \frac{2}{(\mathbf{r}- \mathbf{r}')^2
    [r'^2+(\mathbf{r}- \mathbf{r}')^2]}.
\end{equation}
Inserting this expression in Eq.~(\ref{eq:bfkl1}) we see that the singularity
in the gain and loss terms exactly cancel when $\mathbf{r}= \mathbf{r}'$.
Making also the variable
transformation $\mathbf{k} = \mathbf{r}/r^2$, we arrive at a conventional form
for the BFKL equation in momentum space.
The cancellation of the singularity at $\mathbf{r}'-\mathbf{r}=0$ then 
corresponds to the soft cancellation when $\mathbf{k}_\perp'-\mathbf{k}_\perp=0$ 
in momentum space.

To understand the qualitative features of BFKL evolution we approximate the dipole
distribution in the gain term by its asymptotic form for small and large $r$:
\begin{equation}
\frac{r'^2}{r^2(\mathbf{r}- \mathbf{r}')^2}\left\{ 
\begin{array}{ll}
\approx \frac{1}{r^2} \,\,\,\mathrm{for}\,\,\,r<r'\\ 
\approx \frac{r'^2}{r^4} \,\,\,\mathrm{for}\,\,\,r>r'
\end{array}
\right.
\end{equation}
This approximation is non-singular when $\mathbf{r'}-\mathbf{r}\rightarrow 0$. 
In Eq.~(\ref{eq:bfkl1}) the singularity in this point was canceled by the loss 
term, and in this approximation we therefore now only keep the gain term.
The result is the equation
\begin{equation}
\frac{\partial \mathcal{F}(Y, r^2)}{\partial Y} \approx\bar{\alpha} 
\left\{ \int_{r^2} \frac{d r'^2}{r^2} \mathcal{F}(Y,r'^2) +
\int^{r^2} \frac{d r'^2 r'^2}{r^4} \mathcal{F}(Y,r'^2) \right\}
\label{eq:bfklappr}
\end{equation}

We now make the ansatz
\begin{equation}
\mathcal{F}(Y, r^2) \sim \e^{\lambda Y} (r^2)^{-\gamma -1},
\label{eq:ansatz}
\end{equation}
which inserted in Eq.~(\ref{eq:bfklappr}) gives
\begin{equation}
\lambda\mathcal{F}=\bar{\alpha}\left[\frac{1}{\gamma} +
  \frac{1}{1-\gamma}\right] \mathcal{F}
\label{eq:gammaappr}
\end{equation}
This approximation reproduces the qualitative
features of the LL BFKL equation, with singularities at $\gamma=0$ and 
$\gamma=1$. The right hand side has an extreme point for $\gamma=0.5$, which 
corresponds to $\lambda=4\bar{\alpha}$. The approximation somewhat
overestimates the contribution from the region $r'\approx r$, and therefore
the $\lambda$-value is larger than the true value $\lambda=4\ln2\,
\bar{\alpha}$. The solution corresponds to an exponential growth for large
$Y$, $\sim \e^{\lambda Y}\sim 1/x^\lambda$, which is thus faster than the DLL
result given by the exponential of the square root of $Y$.


The BFKL equation describes the density of partons in a cascade, which is
relevant for \emph{inclusive} cross sections. 
\emph{Exclusive final states} can be calculated in the CCFM model 
\cite{Catani:1989sg, Ciafaloni:1987ur}, which
reproduces BFKL evolution in terms of weights for final states, in
which all real gluons are ordered in angle and rapidity. This will be further
discussed in Sec.~\ref{sec:lunddipole}.

\section{Multiple interactions and saturation}
\label{sec:sat}

\subsection{Experimental evidence}

\begin{figure}
\begin{center}
\epsfig{file=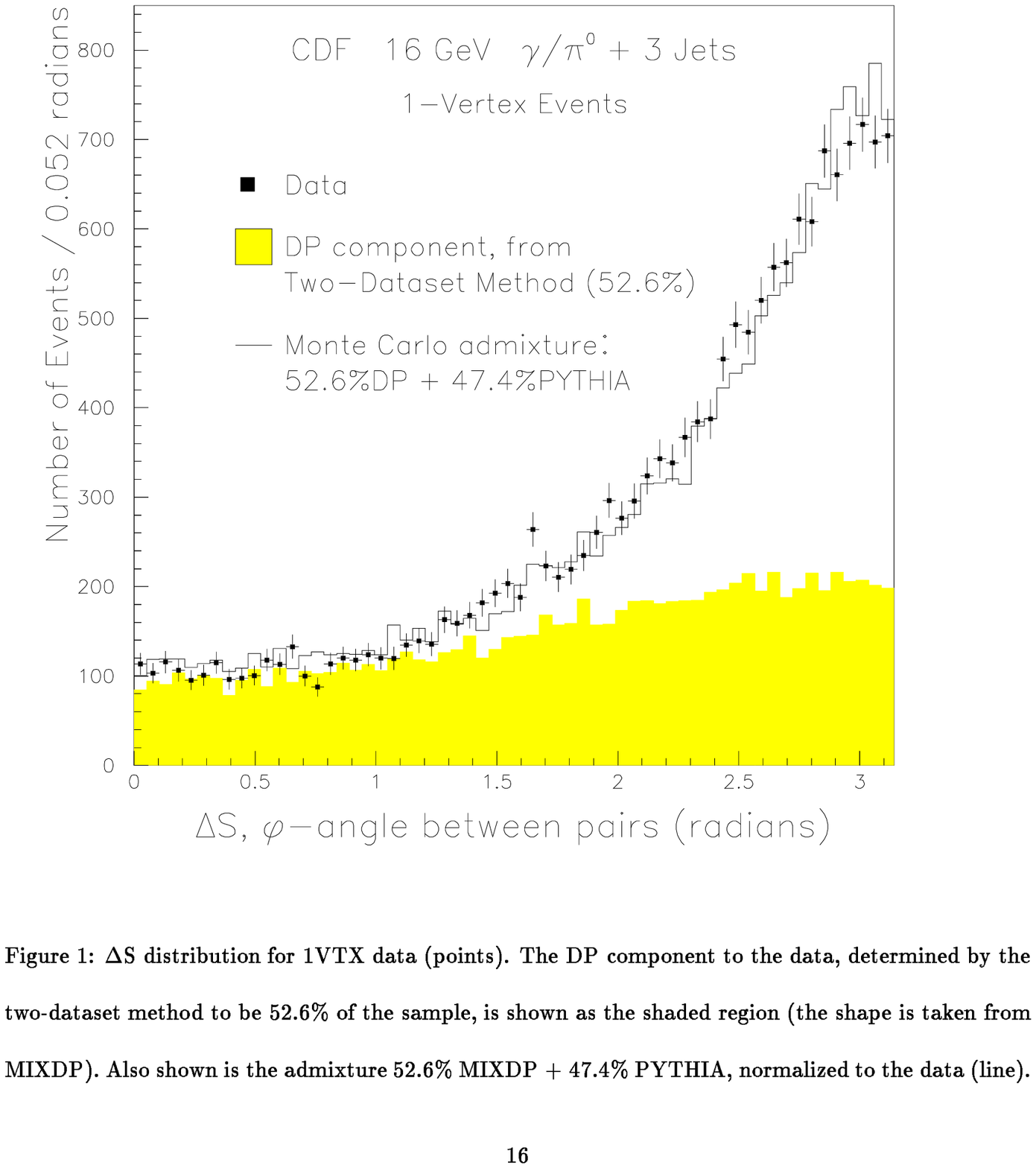,width=75mm,%
clip=, bbllx=100,bblly=180,bburx=550,bbury=650}
\end{center}
\label{fig:cdfmi}
\caption{Distribution in azimuth angle between pairs in events with
  $\gamma/\pi^0 +$ 3 jets from the CDF coll. \cite{Abe:1997xk}. 
The shaded (yellow) region shows expectation from double parton scattering.}
\end{figure}

The strong increase in the parton density at high energy implies that a single
event often contains multiple parton-parton subcollisions. Such events have
been observed experimentally \cite{Akesson:1986iv, Abe:1993rv, Abe:1997xk, 
Abazov:2002mr}. As an example Fig.~\ref{fig:cdfmi} shows 
results from CDF for events with 3 jets + $\gamma$, which can only be
described including multiple hard subcollisions. 
\emph{Cf} also the talk by Rick Field at this school \cite{arXiv:1110.5530}

\subsection{Eikonal formalism}

As mentioned above, rescattering and multiple interactions are most easily 
treated in impact parameter space. The result of repeated scattering with
momenta $\mathbf{k}_{\perp i}$ is given by a convolution in
$\mathbf{k}_\perp$-space, which corresponds to a multiplication in
$\mathbf{b}$-space. Thus in impact parameter space the multiple interactions 
are described
by a product of the $S$-matrix elements for the individual interactions:
\begin{equation}
S(b)=S_1(b) S_2(b) S_3(b).
\end{equation}

If the interaction is driven by absorption into inelastic states $i$, 
with weights $2f_i$, the optical theorem gives an elastic amplitude given by
\begin{equation}
T = 1-e^{-F},\,\,\,\,\,\mathrm{with}\,\,\,F=\sum f_i.
\end{equation}
For a structureless projectile we then find
\vspace{3mm}
\begin{equation}
\left\{ \begin{array}{l}
d\sigma_{\mathrm{tot}} / d^2b =\,\langle 2T \rangle,\,\,\,\,\\
\sigma_{\mathrm{el}}/d^2b \,\,=\, \langle T\rangle^2,\,\,\,\,\\
\sigma_{\mathrm{inel}}/d^2b = \,\langle 1-e^{-\sum 2f_i}\rangle = \sigma_{\mathrm{tot}}-
\sigma_{\mathrm{el}}.
\end{array} \right.
\end{equation} 

\subsection{Diffractive excitation, Good--Walker formalism}
\label{sec:GW}

If the projectile has an internal structure, the mass eigenstates $\Psi_{k}$
can differ from the eigenstates of diffraction $\Phi_n$, which have
eigenvalues $T_n$. With the notation
$\Psi_{k} = \sum_n  c_{kn} \Phi_n\,\,\,(\mathrm{with}\,\,\,\Psi_{in}=\Psi_1)$
the elastic amplitude is given by
$\langle \Psi_{1} | T | \Psi_{1} \rangle = \sum c_{1n}^2 T_n
= \langle T \rangle$,
while the amplitude for diffractive transition to mass eigenstate $\Psi_k$ is 
given by
$\langle \Psi_{k} | T | \Psi_{1} \rangle = \sum_n  c_{kn} T_n c_{1n}$.
The corresponding cross sections become
\begin{eqnarray}
d \sigma_{\mathrm{el}}/d^2 b &=& (\sum c_{1n}^2 T_n)^2 = \langle T\rangle ^2\\
d\sigma_{\mathrm{diff}}/d^2 b &=&\sum_k \langle \Psi_{1} | T | \Psi_{k} \rangle \langle 
\Psi_{k} | T |\Psi_{1} \rangle =\langle T^2 \rangle.
\end{eqnarray}
The diffractive cross section here includes elastic scattering. Subtracting
this gives the cross section for diffractive excitation, which is thus 
determined by the fluctuations in the scattering process:
\begin{equation}
d\sigma_{\mathrm{diff\,ex}}  = d\sigma_{\mathrm{diff}}- d \sigma_{\mathrm{el}} =
(\langle T^2 \rangle - \langle T \rangle ^2)d^2 b.
\end{equation}

\subsection{The BK equation and saturation}

Consider scattering of a dipole with charges at transverse coordinates
$\mathbf{x}$ and $\mathbf{y}$ against a dense target at rapidity distance
$Y$. The interaction probability is called $N(\mathbf{x, y}, Y)$. Study the
change in interaction probability when $Y$ is changed to $Y+\delta Y$. The
probability that the dipole has emitted a gluon at point $\mathbf{z}$,
within the interval $\delta Y$,
is given by Eq.~(\ref{eq:transversedipole}). The change in interaction
probability is therefore given by \cite{Kovchegov:1999ua}
\begin{eqnarray}
\frac{d\,N(\mathbf{x, y}, Y)}{d\,Y}=\frac{\bar{\alpha}}{2\pi} 
\int d^2z \frac{(\mathbf{x- y})^2}{(\mathbf{x- z})^2(\mathbf{z- y})^2}\times
\nonumber \\
\left[ N(\mathbf{x, z},Y)+N(\mathbf{z, y},Y)-N(\mathbf{x, y},Y)-
N(\mathbf{x, z},Y)N(\mathbf{z, y},Y) \right].
\label{eq:BK}
\end{eqnarray}
Here the first two terms in the square bracket give the probability for the new
dipoles to interact, the third term is the reduction because the original
dipole has disappeared, and the last term avoids double counting by
subtracting the probability that both new dipoles interact. This non-linear
term prevents the interaction probability to grow beyond 1.

If we now take the average, and furthermore assume that $\langle N\cdot
N\rangle = \langle N\rangle^2$ (which may be allowed for a sufficiently dense 
and homogenous target), we arrive at the Balitsky-Kovchegov
equation \cite{Kovchegov:1999ua}. 
It is obvious that this equation has two fixpoints, given by $N=0$
and $N=1$. The first corresponds to the weak interaction limit, where the
quadratic term can be neglected. The value $N=1$ corresponds to the black disk
limit, where the interaction probability saturates at the unitarity limit.

\section{Dipole cascade models for high energy collisions}
\label{sec:dip}

\subsection{Mueller's dipole cascade model} 

Mueller's model is based on the dipole evolution discussed in
Secs.~\ref{sec:dipcasc} and \ref{sec:bfkl}, which describes
LL BFKL evolution in transverse coordinate space 
\cite{Mueller:1993rr,Mueller:1994jq,Mueller:1994gb}. 
When a dipole emits a gluon it splits in two dipoles, which in the large $N_c$
limit emit softer gluons independently. The result is a gluon cascade in form
of a dipole chain, as illustrated in Fig.~\ref{fig:dipolesplit}b, where the
number of links grows exponentially with rapidity as discussed in
Sec.~\ref{sec:bfkl}.  
Gluon radiation from the colour charge in a parent quark or gluon is screened 
by the accompanying anticharge 
in the colour dipole, which suppresses emissions at large transverse separation.
Therefore the dipoles become on average smaller and smaller as the cascade
proceeds to smaller rapidities. 

\begin{figure}
\begin{center}
\scalebox{1.3}{
\begin{picture}(180,80)(0,0)

\Text(27,35)[r]{{\footnotesize $i$}}
\Text(53,37)[l]{{\footnotesize $j$}}
\Text(30,29)[tl]{{\footnotesize 2}}
\Text(30,41)[bl]{{\footnotesize 1}}
\Text(51,44)[br]{{\footnotesize 3}}
\Text(50,31)[tr]{{\footnotesize 4}}
\Vertex(150,32){1}
\Vertex(130,30){1}
\Vertex(130,40){1}
\Vertex(151,43){1}
\Vertex(50,32){1}
\Vertex(30,30){1}
\Vertex(30,40){1}
\Vertex(51,43){1}

\Vertex(10,20){1}
\Vertex(18,70){1}
\Vertex(70,20){1}
\Vertex(60,65){1}
\Vertex(20,50){1}
\Vertex(28,57){1}
\Vertex(60,10){1}
\Vertex(60,52){1}

\Vertex(110,20){1}
\Vertex(118,70){1}
\Vertex(170,20){1}
\Vertex(160,65){1}
\Vertex(120,50){1}
\Vertex(128,57){1}
\Vertex(160,10){1}
\Vertex(160,52){1}
\LongArrowArcn(90,20)(20,120,60)

\Line(10,20)(30,30)
\ArrowLine(30,40)(30,30)
\Line(30,40)(20,50)
\Line(20,50)(28,57)
\Line(28,57)(18,70)

\Line(70,20)(60,10)
\Line(60,10)(50,32)
\ArrowLine(50,32)(51,43)
\Line(51,43)(60,52)
\Line(60,52)(60,65)

\ArrowLine(130,30)(110,20)
\ArrowLine(120,50)(130,40)
\Line(120,50)(128,57)
\Line(128,57)(118,70)

\Line(170,20)(160,10)
\ArrowLine(160,10)(150,32)
\ArrowLine(151,43)(160,52)
\Line(160,52)(160,65)

\ArrowLine(150,32)(130,30)
\ArrowLine(130,40)(151,43)

\end{picture}
}
\end{center}
\caption{An interaction between a dipole in the projectile and another in the
  target due to gluon exchange gives a recoupling of the dipole chains.}
\label{fig:dipscatt}
\end{figure}
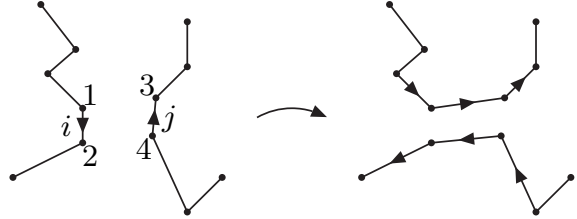

When two cascades collide, a pair of dipoles with coordinates 
$(\mathbf{x}_i,\mathbf{y}_i)$ and $(\mathbf{x}_j,\mathbf{y}_j)$ can interact 
via gluon exchange with the probability $2f_{ij}$, where
\begin{equation}
  f_{ij} = f(\mathbf{x}_i,\mathbf{y}_i|\mathbf{x}_j,\mathbf{y}_j) =
  \frac{\alpha_s^2}{8}\biggl[\log\biggl(\frac{(\mathbf{x}_i-\mathbf{y}_j)^2
    (\mathbf{y}_i-\mathbf{x}_j)^2}
  {(\mathbf{x}_i-\mathbf{x}_j)^2(\mathbf{y}_i-\mathbf{y}_j)^2}\biggr)\biggr]^2.
\label{eq:dipamp}
\end{equation}
We note here in particular that the interaction 
probability goes to zero for a small dipole. This implies that the singularity
in the production probability for small dipoles in 
Eq.~(\ref{eq:transversedipole}) does not give infinite cross sections.
We note also that gluon exchange means exchange of colour between the two 
cascades. This implies a reconnection of the dipole chains, as shown in
Fig.~\ref{fig:dipscatt}, and the formation of dipole chains connecting the
projectile and target remnants. 

\begin{figure}
\begin{center}
\includegraphics[width=.85\linewidth]{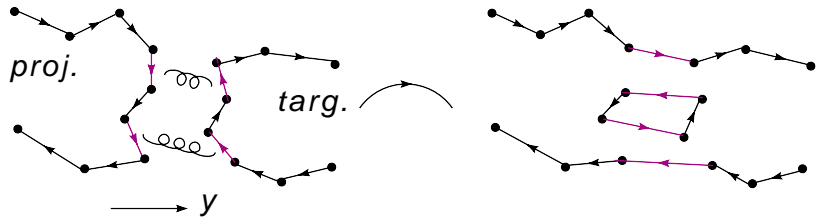}
\end{center}
\caption{Double interaction results in a dipole loop, corresponding to a pomeron
  loop.}
\label{fig:loop}
\end{figure}

In Mueller's model the constraints from unitarity are satisfied using the 
eikonal formalism.
When more than one pair of dipoles interact, colour loops are formed, as shown
in Fig.~\ref{fig:loop}. 
This double interaction is an effect of saturation, corresponding to the 
non-linear term in the BK equation
(\ref{eq:BK}). It is also related to multiple pomeron exchange and pomeron
loops in the regge formalism.

In the schematic illustration in Fig.~\ref{fig:loop}, rapidity is growing along
the horizontal direction. We here note that if this event was analysed in a
Lorentz frame closer to the target, the dipole loop could lie completely
within the evolution of the projectile. Thus double interaction in one frame can
correspond to a colour loop within the evolution, when viewed in a different 
Lorentz frame. Such loops are not included in Mueller's model, and are also
not taken into account in the BK equation.

\subsection{Lund dipole cascade model}
\label{sec:lunddipole}

The Lund model \cite{Avsar:2005iz, Avsar:2006jy,
  Flensburg:2008ag, Flensburg:2011kk} is a 
generalization of Mueller's model, which also includes:

--  NLL BFKL effects

-- Nonlinear effects in the evolution

-- Confinement effects

It is implemented in a MC called DIPSY, with applications to collisions
between electrons, protons, and nuclei. An incoming virtual photon is here
treated as a $q\bar{q}$ pair, with an initial state wavefunction determined
by QED.  For an incoming proton we make an ansatz in form of an
equilateral triangle of dipoles, but after evolution the result is rather
insensitive to the exact form of the initial state.
\vspace{2mm}

\subsubsection{Beyond LL BFKL effects}

The NLL corrections to BFKL evolution have three major sources
\cite{Salam:1999cn}:

\emph {Non-singular terms in the splitting function:}
These terms suppress large $z$-values in the individual parton branchings. 
Most of this effect is taken care of by including energy-momentum
conservation. This is effectively taken into account by associating a
dipole with transverse size $r$ with a transverse momentum $k_\perp = 1/r$,
and demanding conservation of the lightcone momentum $p_+$ in every step
in the evolution. This gives an effective cutoff for small dipoles.

\emph {Projectile-target symmetry:}
A parton chain should look the same if generated from the target end
as from the projectile end. The corresponding corrections are also
called energy scale terms, and are essentially equivalent to the so
called consistency constraint \cite{Kwiecinski:1996td}. This effect is
taken into account by conservation of the negative
lightcone momentum components, $p_-$.

\emph {The running coupling:}
Following Ref.~\cite{JLAB-THY-07-679}, the scale in the running coupling is
taken as the largest transverse momentum in the vertex.
\vspace{2mm}

\subsubsection{Nonlinear effects in the evolution}

\begin{figure}
\begin{center}
\scalebox{0.95}{\mbox{
\begin{picture}(180,100)(0,0)
\Line(10,10)(20,25)
\Line(25,40)(20,25)
\Line(50,40)(80,30)
\Line(75,60)(80,30)
\Line(75,60)(55,50)
\Line(30,50)(15,70)
\Line(15,70)(10,90)
\LongArrow(85,45)(100,45)
\Line(100,10)(110,25)
\Line(115,40)(110,25)

\Line(140,40)(170,30)
\Line(165,60)(170,30)
\Line(165,60)(145,50)

\Line(120,50)(105,70)
\Line(105,70)(100,90)
\Vertex(10,10){2}
\Vertex(20,25){2}
\Vertex(25,40){2} 
\Vertex(50,40){2} 
\Vertex(80,30){2} 
\Vertex(75,60){2}
\Vertex(55,50){2} \Vertex(30,50){2} \Vertex(15,70){2} \Vertex(10,90){2}
\Vertex(100,10){2} \Vertex(110,25){2} \Vertex(115,40){2} \Vertex(120,50){2}
\Vertex(140,40){2} \Vertex(170,30){2} \Vertex(165,60){2} \Vertex(145,50){2}
\Vertex(105,70){2} \Vertex(100,90){2}

\Line(25,40)(50,40)
\Line(30,50)(55,50)
\Line(115,40)(120,50)
\Line(145,50)(140,40)
\Text(25,39)[tl]{$\bar{r}$}
\Text(50,37)[tr]{$r$}
\Text(30,53)[bl]{$r$}
\Text(55,52)[br]{$\bar{r}$} 
\end{picture}
}}
\end{center}
\caption{Two dipoles with the same colour form a colour octet, which may be
  better approximated by dipoles formed by the closet colour-anticolour
  pairs. This implies a recoupling of the dipole chains.}
\label{fig:swing}
\end{figure}
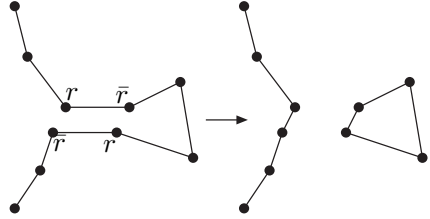

As mentioned above, multiple interactions produce loops of dipole chains
corresponding to pomeron loops. Mueller's model includes all loops cut in 
the particular Lorentz frame used for the analysis, but not loops contained
within the evolution of the individual projectile and target cascades.  
As for dipole scattering
the probability for such loops is given by $\alpha_s$, and therefore
formally colour suppressed compared to dipole splitting, which is
proportional to $\bar{\alpha}=N_c \alpha_s/\pi$. These loops are therefore
related to the probability that two dipoles have the same colour. Two dipoles
with the same colour form a quadrupole field. Such a field may be better
approximated by two dipoles formed by the closest colour--anticolour
charges. This corresponds to a recoupling of the colour dipole chains. The
process is illustrated in Fig.~\ref{fig:swing}, and we call
it a dipole ``swing''.  With a weight for the swing which favours small
dipoles, we obtain an almost frame independent result.
The number of dipoles in the cascade is not reduced, and the saturation 
effect is a consequence of the smaller interaction probability for 
the smaller dipoles. Thus the number of \emph{interacting} dipoles is
reduced. Counting only these ``effective'' dipoles, the swing can be looked
upon as a $2\rightarrow 1$, or in some cases $2\rightarrow 0$, transition.

\vspace{2mm}

\subsubsection{Confinement} 

Confinement is also important. A purely perturbative evolution with massless
gluons violates Froissart's bound \cite{Avsar:2008dn}. This is avoided by 
giving the gluon an effective mass.

\subsection{Results}

\subsubsection{Total and elastic cross sections}

\begin{figure}
\raisebox{44mm}{\includegraphics[width=0.33\linewidth,angle=270]{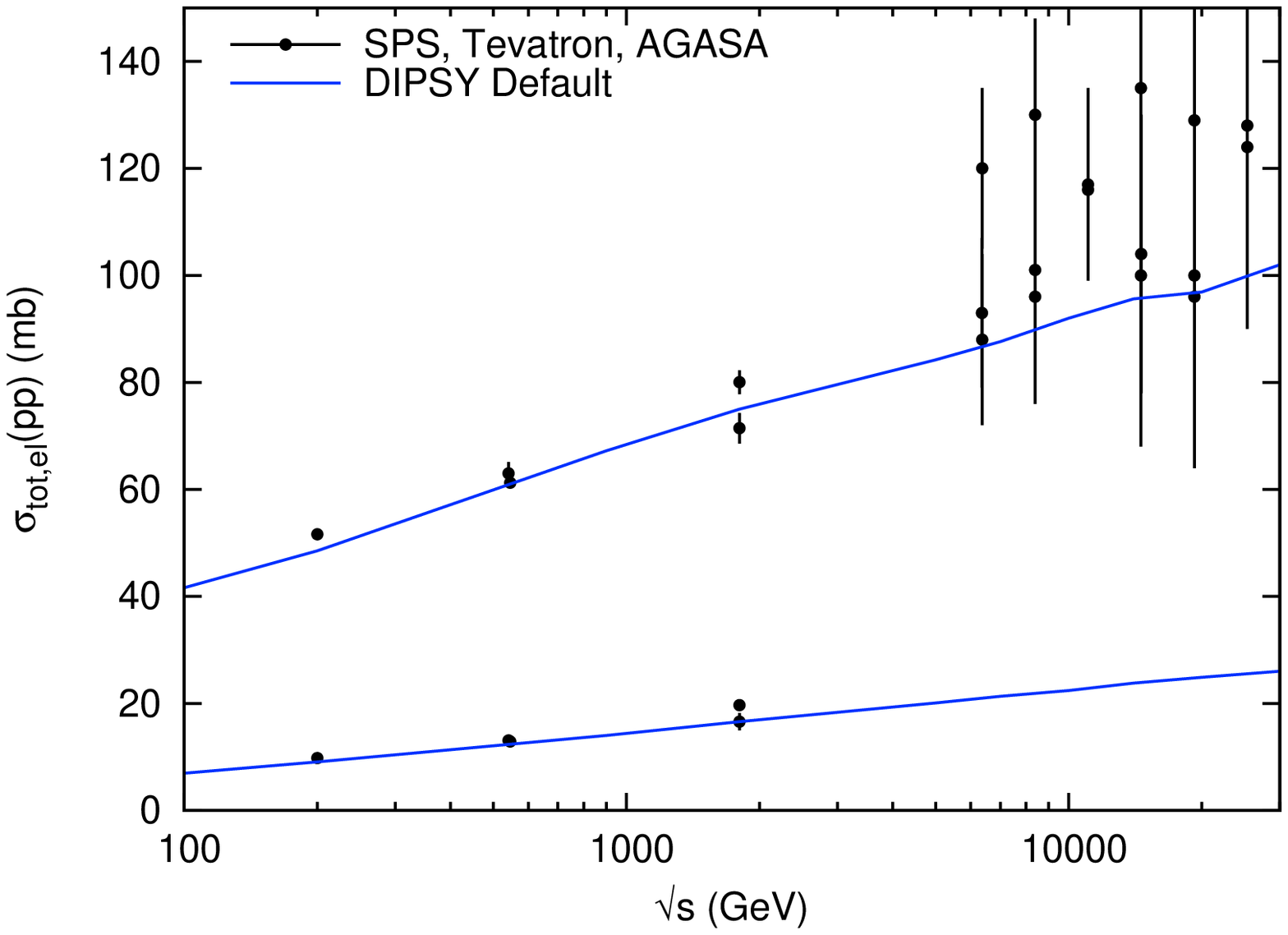}}
\includegraphics[width=0.53\linewidth,angle=0]{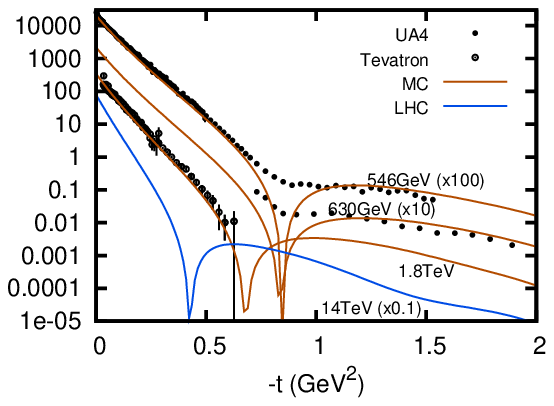}
\caption{\label{fig:pptot}Total and elastic cross sections in $pp$
  collisions in the dipole cascade model. }
\end{figure}

\begin{figure}
\raisebox{41mm}{\includegraphics[width=0.33\linewidth,angle=270]{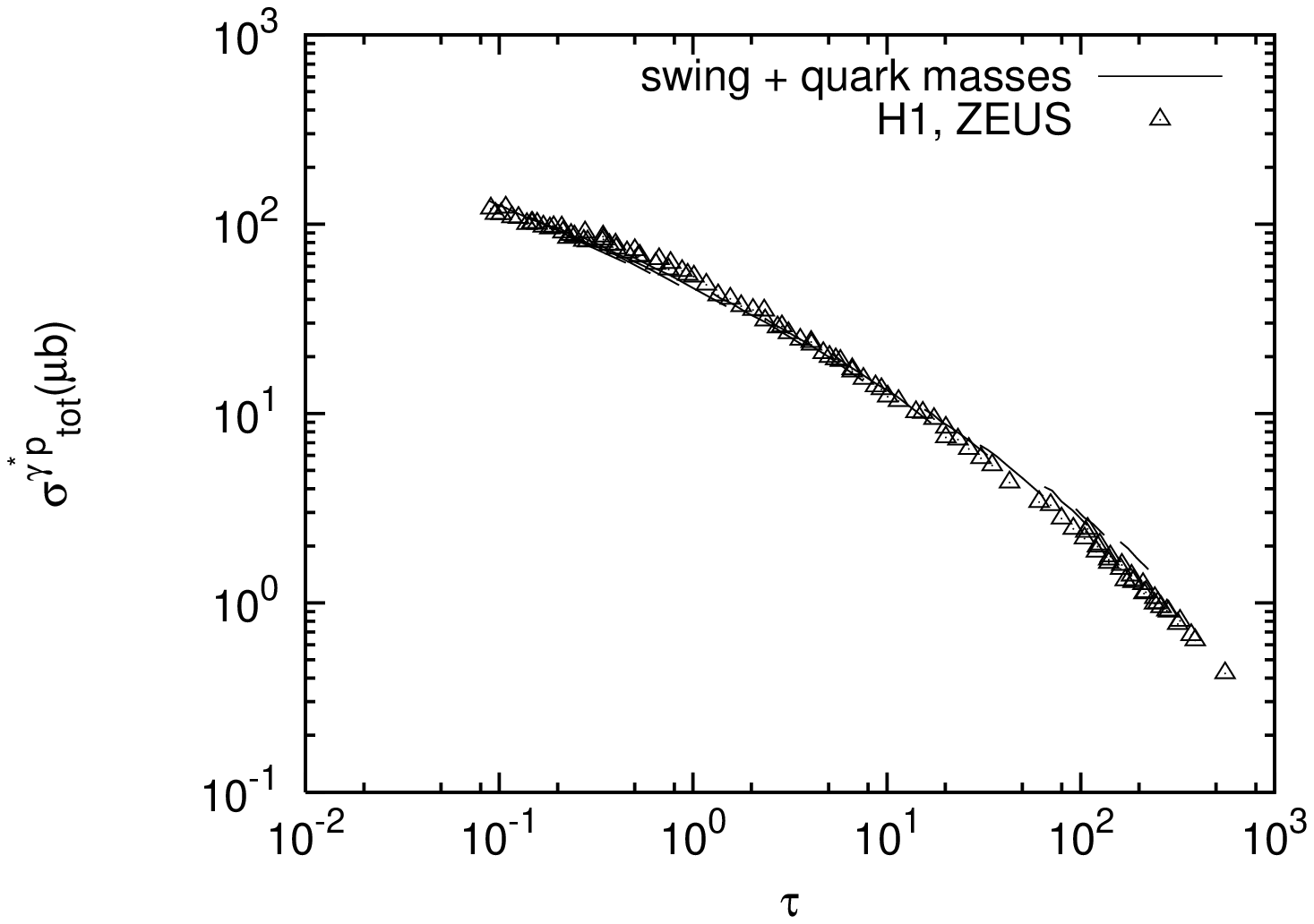}}
 \includegraphics*[bb=50 52 230 160, width=0.53\linewidth]{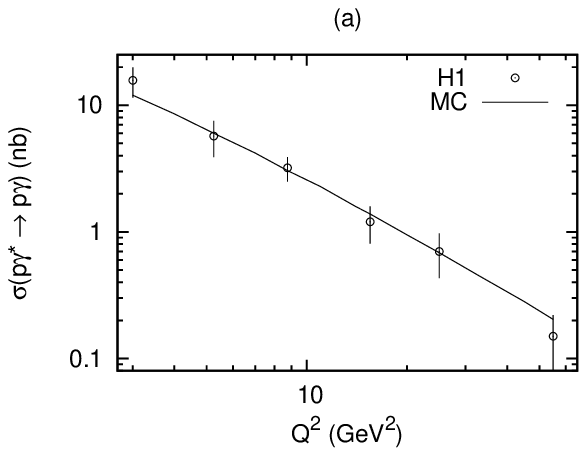}
\caption{\label{fig:distot}\emph{Left}: Total $\gamma^*p$ cross section for
  combinations of $x$ and $Q^2$, presented as a
 function of the scaling parameter 
$\tau=(Q^2/Q_0^2)(x/x_0)^\lambda$ \cite{hep-ph/0007192}, with $Q_0=1$GeV, $x_0=3\cdot10^{-4}$, 
and $\lambda=0.29$. \emph{Right}: The cross section for $\gamma^\star p\to
    \gamma p$ (DVCS) for $W = 82$~GeV as function of $Q^2$. Data from H1
    \cite{hep-ex/0505061}.} 
\end{figure}

Results for total and elastic $pp$ cross sections are presented in 
Fig.~\ref{fig:pptot}. Corresponding results for total and quasielastic
scattering in DIS are shown in 
Fig.~\ref{fig:distot}. We here see that the the experimental data are very well
reproduced by the model.

\subsubsection{Diffractive excitation}

Diffractive excitation accounts for large fractions of the cross sections in
DIS and $pp$ collisions. 
As mentioned in Sec.~\ref{sec:GW}, diffractive excitation is in the
Good--Walker formalism determined by the fluctuations in the scattering 
amplitude, and we note that the BFKL evolution gives large fluctuations in 
the cascade evolution. 

\begin{figure}
\begin{center}
\includegraphics[width=0.47\linewidth,angle=0]{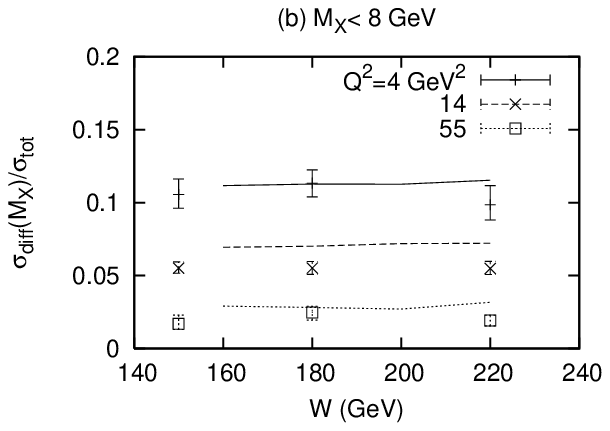}
\raisebox{40mm}{\includegraphics[width=0.35\linewidth,angle=270]{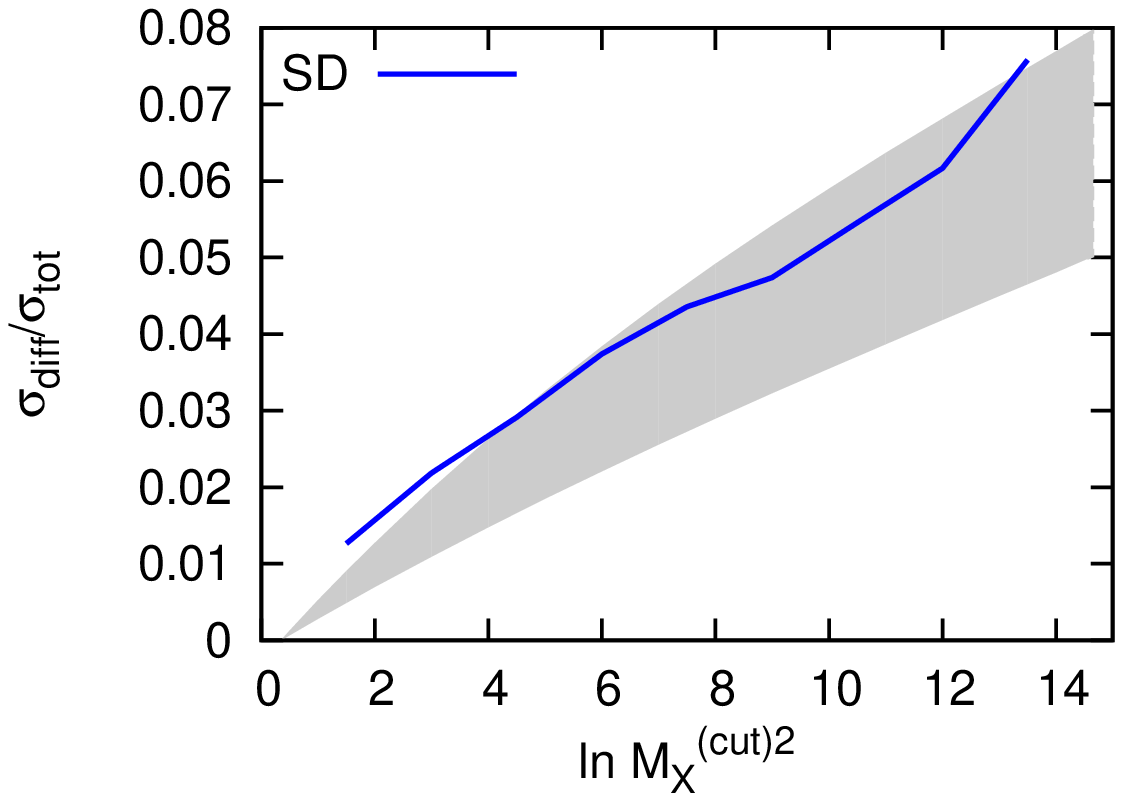}}

\end{center}
\caption{\emph{Left}: The ratio of the total diffractive
    cross section to the total cross section in DIS as a function of $W$, for
    $M_X< 8$ GeV. Data from \textsc{Zeus} \cite{hep-ex/0501060}.
\emph{Right}: The single diffractive cross section for $M_X < M_X^{(cut)}$ in $pp$
  collisions at 1.8 TeV. The
  shaded region is an estimate from CDF data \cite{FERMILAB-PUB-93-233-E}.}
\label{fig:diff}
\end{figure}

Study the interaction in a frame, where the projectile is evolved a 
distance $y_1$ and the target $y_2=Y-y_1$, with $Y$ the total rapidity
range $\approx \ln s/(1 \textrm{GeV}^2)$. 
If we here first take the average over the target states, we get
the amplitude for elastic scattering of the target. Squaring it gives the
cross section, when the target is scattered elastically. If we after this take 
the average over the projectile states, we obtain the diffractive scattering
of the projectile, including the elastic scattering. Thus the expression
\begin{equation}
\langle \langle T \rangle_{targ}^2 \rangle_{proj} -  
\langle T \rangle_{targ,proj}^2 
\end{equation}
gives the cross section for single diffractive excitation of the projectile,
with the excited mass limited to $M_X^2 < \exp(y_1)$. Varying $y_1$ gives 
then $d\sigma/d M_X^2$. The
resulting cross sections for diffractive excitation in DIS and $pp$ collisions
are shown in Fig.~\ref{fig:diff}, together with \textsc{Zeus} data for DIS,
and an estimate from CDF data for $p\bar{p}$ collisions.

\begin{figure}
\includegraphics[scale=0.44,angle=270]{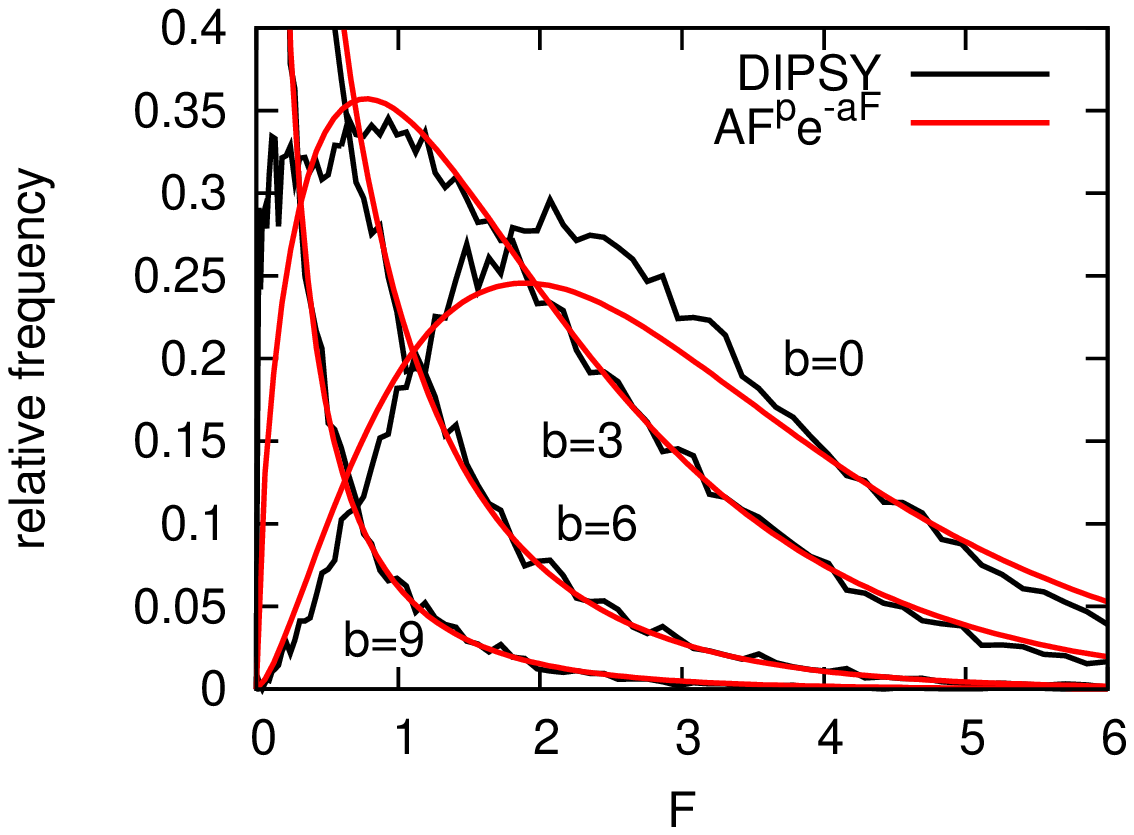}
\includegraphics[scale=0.44,angle=270]{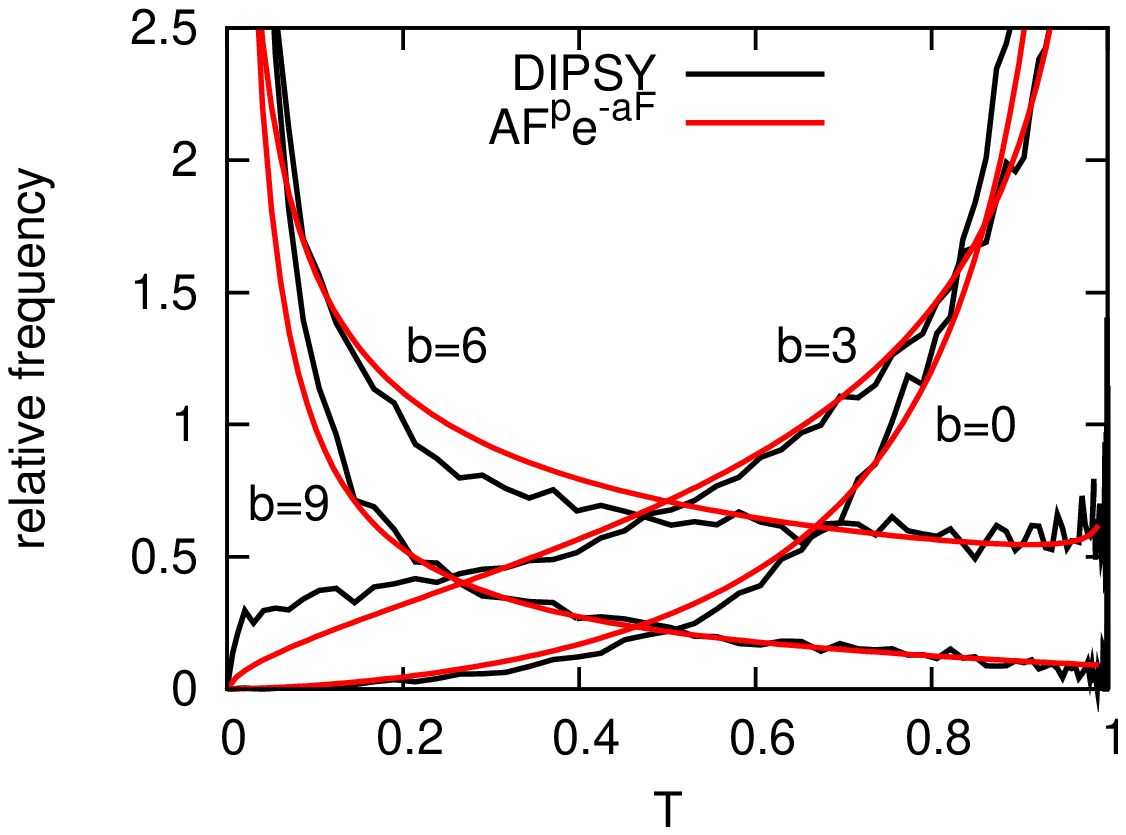}
\caption{Distribution in the one-pomeron
   amplitude $F$ (left), and the uniterized amplitude $T$ (right)
    in $pp$ collisions at 2 TeV. $b$ is in units of GeV$^{-1}$.}
\label{fig:ftdistpp}
\end{figure}

It is interesting to study the effects of saturation on diffractive excitation
\cite{arXiv:1004.5502}. Saturation is not very important in DIS, but
in $pp$ scattering the Born amplitude is large, and therefore the unitarity
effects are also large. Fig.~\ref{fig:ftdistpp} shows both the Born amplitude 
and the unitarized amplitude at 2 TeV for different impact parameters $b$. We 
see that the width of the Born amplitude is very large, and without 
unitarization the
fraction of diffractive excitation would be correspondingly large. 
(The smooth lines are fits of the form $A F^p e^{-aF}$.)

However, the unitarized amplitude is limited
by 1, and the width of the distribution, and therefore the diffractive 
excitation, is very much reduced. This result corresponds to 
the  effect of enhanced diagrams in the conventional triple-regge approach.
We note here also that the strong suppression from saturation implies that
factorization is broken when comparing diffraction in $pp$ collisions and DIS
\cite{hep-ex/0209001}.

\begin{figure}
\begin{center}
\includegraphics[scale=0.38,angle=270]{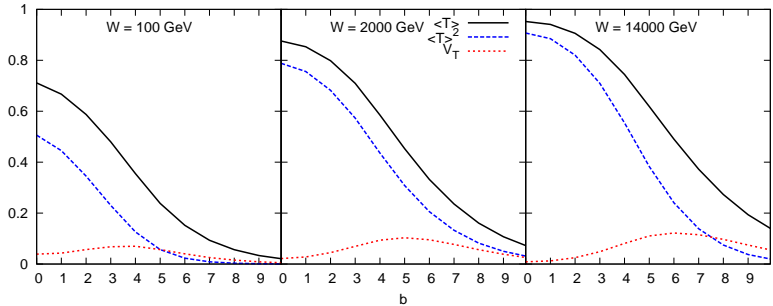}
\caption{\label{fig:profile}Impact parameter distributions for 
    $\langle T\rangle=(d\sigma_{\mathrm{tot}}/d^2b)/2$, 
    $\langle T\rangle^2=d\sigma_{\mathrm{el}}/d^2b$, and 
    $V_T=d\sigma_{\mathrm{diff\, ex}}/d^2b$ in
    $pp$ collisions at $W=2$ TeV. $b$ is in units of GeV$^{-1}$.}
\end{center}
\end{figure}

The absorption is most important for central collisions, where thus
diffractive excitation is most strongly suppressed. As shown in
Fig.~\ref{fig:profile}, the cross section for diffractive
excitation is therefore largest in a ring with radius 
$b\sim 1\, \mathrm{fm}\approx 5\,\mathrm{GeV}^{-1}$, which
grows slowly with energy.

\subsubsection{Comparison with Multi-Regge Analyses}

It is also interesting to compare the results from the Good--Walker analysis
with the multi-regge formalism. To this end we study the contribution from the
\emph{bare pomeron}, meaning the one-pomeron amplitude without contributions 
from saturation, enhanced diagrams or gap survival form factors.

When $s$, $M_{\mathrm{X}}^2$, and $s/M_{\mathrm{X}}^2$ are all large, pomeron exchange should 
dominate. If the pomeron is a simple pole, we expect the following 
expressions for the $pp$ total and diffractive cross sections:
\begin{eqnarray}
\sigma_{\mathrm{tot}}&=&\beta^2(0)(s/s_0)^{\alpha(0)-1} =\beta^2(0)(s/s_0)^\epsilon,\nonumber\\
\frac{d\sigma_{\mathrm{el}}}{dt} &=& \frac{1}{16\pi}\beta^4(t)(s/s_0)^{2(\alpha(t)-1)}, \nonumber\\
M_{\mathrm{X}}^2 \frac{d\sigma_{\mathrm{SD}}}{dtd(M_{\mathrm{X}}^2)} &=& \frac{1}{16\pi}\beta^2(t)\beta(0)g_{3\mathrm{P}}(t)
\left( \frac{s}{M_{\mathrm{X}}^2} \right)^{2(\alpha(t)-1)}\left( M_{\mathrm{X}}^2 \right)^{\epsilon}.
\label{eq:barepomeron}
\end{eqnarray}
Here $\alpha(t)=1+\epsilon+\alpha' t$ is the 
pomeron trajectory, and $\beta(t)$ and $g_{3P}(t)$ are the proton-pomeron and
triple-pomeron couplings respectively. 
Comparing our result with this expression we find that it indeed reproduces
the triple pomeron form, with the following parameter values obtained choosing
the value $s_0=1\mathrm{GeV}^2$ for the arbitrary scale parameter 
\cite{arXiv:1004.5502}:
\begin{eqnarray}
\alpha(0)&=&1+\epsilon =1.21, \,\,\,\,\alpha' = 0.2\,\mathrm{GeV}^{-2},\nonumber\\
\beta^2(0)&=&12.6\,\mathrm{mb}, \,\,\,\,
\beta(t)=\beta(0)\exp\left(\frac{2.5\,t}{1-1.8\,t}\right),\nonumber\\
 g_{3\mathrm{P}}(t)&=&\mathrm{const.} = 0.3\,\mathrm{GeV}^{-1}.
\label{eq:parameters}
\end{eqnarray}

\subsubsection{Correlations}

We define the double parton distribution, and the impact parameter profile 
$F$ by the relation 
\begin{equation}
\Gamma(x_1,x_2,b;Q_1^2,Q_2^2) \equiv D(x_1, Q_1^2)\,D(x_2, Q_2^2)
\,F(b;x_1,x_2,Q_1^2,Q_2^2),
\end{equation}
where $D(x,Q^2)$ is the single parton distribution.
This implies that the cross section for double hard interactions at midrapidity 
is given by
\begin{equation}
\sigma^D_{(A,B)} \equiv \frac{1}{(1+\delta_{AB})} \frac{\sigma^S_A
  \sigma^S_B}{\sigma_{\mathrm{eff}}},
\end{equation}
with the ``effective cross section'', $\sigma_{\mathrm{eff}}$,
determined by the relation
\begin{equation}
\sigma_{\mathrm{eff}}  = \left[\int d^2b (F(b))^2 \right]^{-1}.
\end{equation}

\begin{figure}
\includegraphics[width=0.49\linewidth,angle=0]{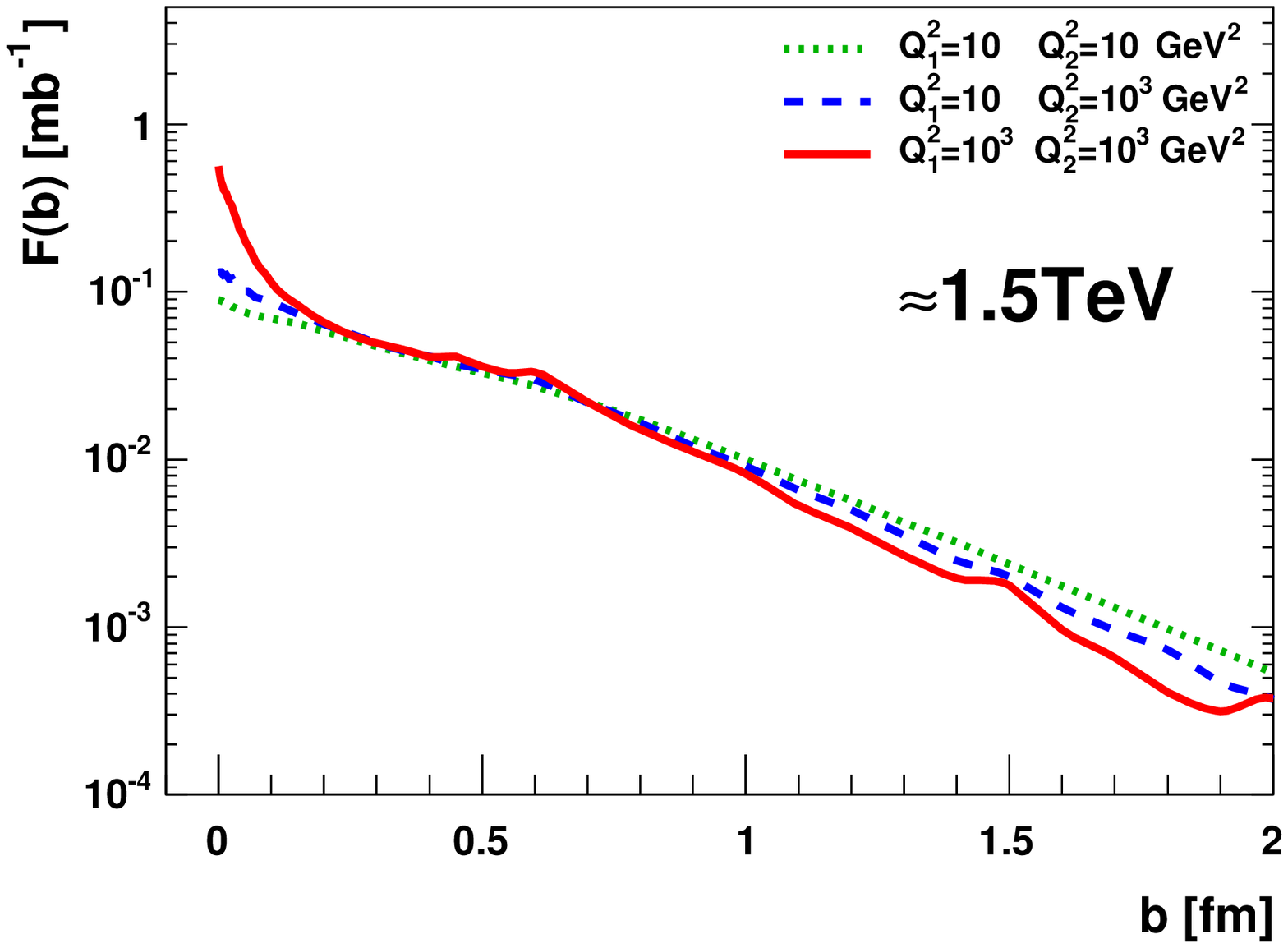}
\includegraphics[width=0.49\linewidth,angle=0]{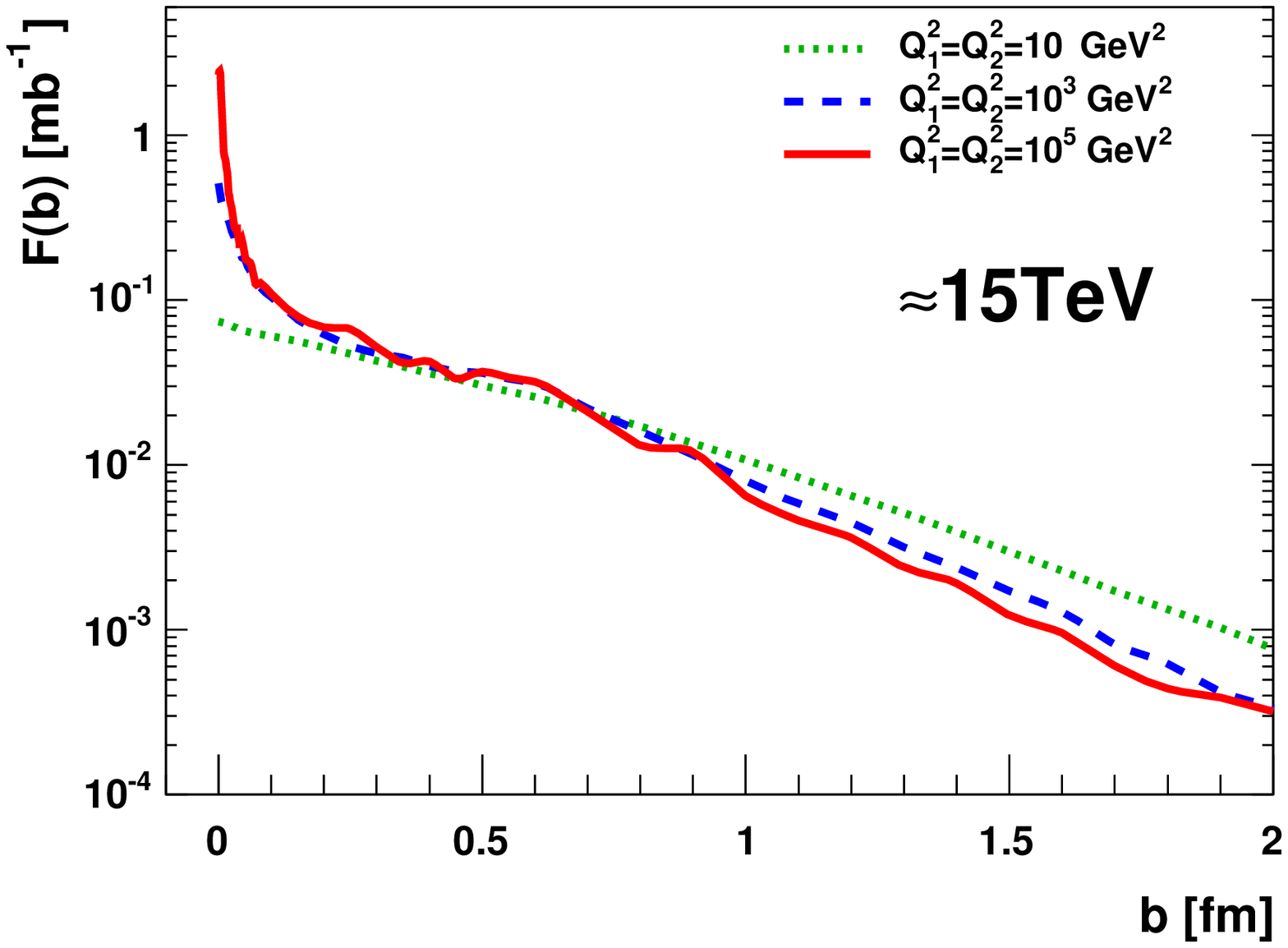}
\caption{Correlation function $F(b)$ for midrapidity subcollisions at different
  $Q^2$ at Tevatron and LHC energies.}
\label{fig:correl}
\end{figure}

$F$ and $\sigma_{\mathrm{eff}}$ are often assumed to depend only
weakly on $x_i$ and $Q_i^2$. The DIPSY MC shows instead that a spike (hotspot)
develops for small separations $b$ at larger $Q^2$, as illustrated in
Fig.~\ref{fig:correl} \cite{Flensburg:2011kj}. This result implies that 
$\sigma_{\mathrm{eff}}$ depends strongly 
on $Q^2$ for fixed $\sqrt s$, as illustrated in Table~\ref{tab:sigmaeff}.
Part of the correlation is due to fluctuations in the cascade, which can also
be taken into account in the MC. Without fluctuations $\int\! F d^2b$ should be
1. In Table~\ref{tab:sigmaeff} we see that the fluctuations increase $\int\! F$
by about 10\%, which thus contributes to the correlations given by $\int\! F^2$.

\begin{table}
\begin{center}
\begin{tabular}{|llll|r|r|}
\hline
\hline
\multicolumn{4}{|c|}{$Q_1^2,\,\,\,Q_2^2\mbox{~[GeV$^2$]},\,x_1,\,\,\,x_2$}  &$\sigma_{\mathrm{eff}}$~[mb]  &$\int\! F$\\
\hline
\multicolumn{4}{|c|}{ 1.5~TeV, midrapidity}  & &  \\
   10       &10      &0.001  &0.001         & 35.3  \hspace{3mm}      & 1.09  \\
  $10^3$   &$10^3$   &0.01   &0.01          & 23.1  \hspace{3mm}      & 1.06  \\
\hline
\multicolumn{4}{|c|}{ 15~TeV, midrapidity}  & &   \\
   10       &10      &0.0001 &0.0001        & 40.4 \hspace{3mm}       & 1.11 \\
  $10^3$   &$10^3$   &0.001  &0.001         & 26.3  \hspace{3mm}  & 1.07  \\
  $10^5$   &$10^5$   &0.01   &0.01          & 19.6 \hspace{3mm}       & 1.03  \\
\hline
\end{tabular}
\end{center}
\caption
{
Summary of results for $\sigma_{\mathrm{eff}}$ and corresponding
integrals of the correlation function $F$.
}
\label{tab:sigmaeff}
\end{table}
 
\subsubsection{Final states}

In order to generate exclusive final states, obtained when two dipole cascades
collide, we first have to determine which dipoles interact and become 
recoupled in the way shown in Fig.~\ref{fig:dipscatt}. We note here that BFKL 
is a stochastic
process, and the interactions between different dipole pairs are
uncorrelated. This implies that the probability for interaction between 
dipoles $i$ and $j$ is given by $1 - e^{-2 f_{ij}}$, where $f_{ij}$ is 
determined by Eq.~(\ref{eq:dipamp}). 

As mentioned above, the BFKL equation describes the density of partons in a 
cascade, which is relevant for \emph{inclusive} cross sections. 
To describe \emph{exclusive final states} it is necessary to take into account
colour coherence and angular ordering as well as soft radiation. The
latter includes also contributions from the $z=1$ singularity in the gluon
splitting function. These effects are taken into account in the CCFM
formalism \cite{Catani:1989sg, Ciafaloni:1987ur}, which also reproduces
the BFKL result for the inclusive cross section. 

\begin{figure}
\begin{center}
\includegraphics[width=0.7\linewidth,angle=0]{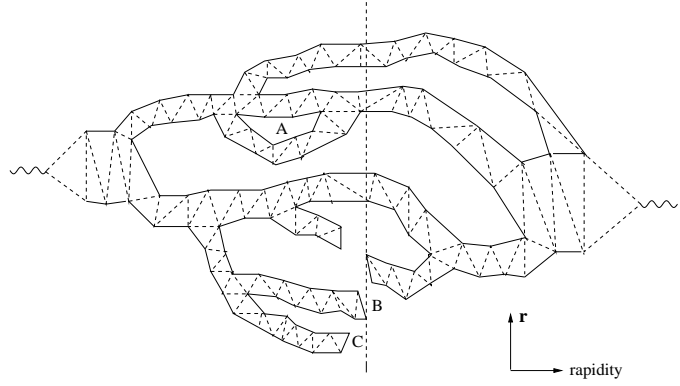}
\end{center}
\caption{Schematic picture of a collision between two dipole cascades. A dipole
loop within the evolution is denoted $A$. Non-interacting branches, like $B$
and $C$ have to be regarded as virtual and reabsorbed.}
\label{fig:multcoll}
\end{figure}

A very schematic picture of a collision between two dipole cascades 
is presented in Fig.~\ref{fig:multcoll}. Here three dipole pairs interact,
forming two dipole loops with an additional loop (denoted $A$) formed within
the evolution of the left cascade. Non-interacting branches, like $B$
and $C$, have to be regarded as virtual and must be reabsorbed.

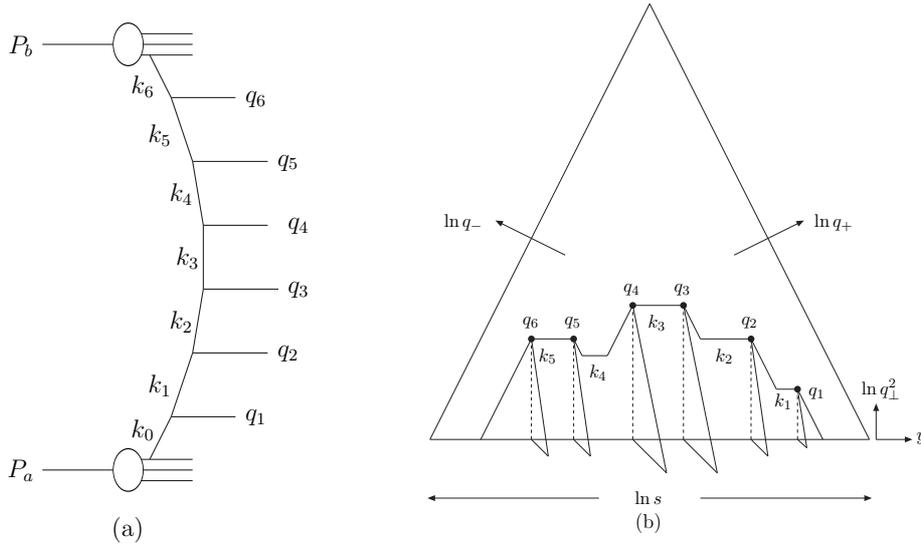
\begin{figure}
\begin{minipage}[]{0.4\linewidth}

\scalebox{0.8}{\mbox{
\begin{picture}(140,260)(-20,-40)
  \Text(-10,15)[]{\large $P_a$}
  \Line(0,15)(40,15)
  \Line(40,20)(70,20)
  \Line(40,15)(70,15)
  \Line(40,10)(70,10)
  \GOval(40,15)(10,7)(0){1}
  \Line(50,20)(60,40)\Text(47,34)[]{\large $k_{0}$}
    \Line(60,40)(90,40)\Text(100,40)[]{\large $q_{1}$}
  \Line(60,40)(70,70)\Text(55,55)[]{\large $k_{1}$}
    \Line(70,70)(105,70)\Text(115,70)[]{\large $q_{2}$}
  \Line(70,70)(75,100)\Text(65,85)[]{\large $k_{2}$}
    \Line(75,100)(110,100)\Text(120,100)[]{\large $q_{3}$}
  \Line(75,100)(75,130)\Text(68,115)[]{\large $k_{3}$}
    \Line(75,130)(105,130)\Text(120,130)[]{\large $q_{4}$}
  \Line(75,130)(70,160)\Text(65,145)[]{\large $k_{4}$}
    \Line(70,160)(105,160)\Text(115,160)[]{\large $q_{5}$}
  \Line(70,160)(60,190)\Text(55,173)[]{\large $k_{5}$}
    \Line(60,190)(90,190)\Text(100,190)[]{\large $q_{6}$}
  \Line(60,190)(50,210)\Text(47,196)[]{\large $k_{6}$}

  \Text(-10,215)[]{\large $P_b$}
  \Line(0,215)(40,215)
  \Line(40,220)(70,220)
  \Line(40,215)(70,215)
  \Line(40,210)(70,210)
  \GOval(40,215)(10,7)(0){1}
  \Text(40,-13)[]{\large (a)}
  
\end{picture}}}
\end{minipage}
\begin{minipage}[]{0.6\linewidth}

\scalebox{0.63}{\mbox{
\begin{picture}(320,240)(15,-23)
  \Line(40,20)(300,20)
  \Line(40,20)(170,280)
  \Line(170,280)(300,20)
  \Line(100,80)(70,20)
  \Text(100,90)[]{$q_{6}$}
  \Vertex(100,80){2}
  \Text(110,70)[]{$k_{5}$}
  \Vertex(125,80){2}
  \Text(125,90)[]{$q_{5}$}
  \Line(100,80)(125,80)
  \Line(125,80)(130,70)
  \Line(130,70)(145,70)
  \Text(140,60)[]{$k_{4}$}
  \Line(145,70)(160,100)
  \Line(160,100)(190,100)
  \Vertex(160,100){2}
  \Text(160,110)[]{$q_{4}$}
  \Text(175,90)[]{$k_{3}$}
  \Vertex(190,100){2}
  \Text(190,110)[]{$q_{3}$}
  \Line(190,100)(200,80)
  \Line(200,80)(230,80)
  \Text(215,70)[]{$k_{2}$}
  \Line(230,80)(245,50)
  \Vertex(230,80){2}
  \Text(230,90)[]{$q_{2}$}
  \Line(245,50)(257.5,50)
 \Text(250,43)[]{$k_{1}$}
  \Vertex(257.5,50){2}
  \Text(270,48)[]{$q_{1}$}
  \Line(257.5,50)(272.5,20)
  \LongArrow(304,20)(304,40)
  \LongArrow(304,20)(324,20)
  \Text(308,50)[]{$\ln q_\perp^2$}
  \Text(332,21)[]{$y$}
  \DashLine(257.5,50)(257.5,20){2}
  \Line(257.5,50)(263,15)
  \Line(257.5,20)(263,15)
  \DashLine(100,80)(100,20){2}
  \Line(100,80)(110,10)
  \Line(100,20)(110,10)
  \DashLine(125,80)(125,20){2}
  \Line(125,80)(135,10)
  \Line(125,20)(135,10)
  \DashLine(160,100)(160,20){2}
  \Line(160,100)(180,0)
  \Line(180,0)(160,20)
  \DashLine(190,100)(190,20){2}
  \Line(190,100)(210,0)
  \Line(210,0)(190,20)
  \DashLine(230,80)(230,20){2}
  \Line(230,80)(240,10)
  \Line(240,10)(230,20)

  \LongArrow(200,-15)(300,-15)
  \LongArrow(140,-15)(40,-15)
  \Text(170,-15)[]{$\ln s$}
  \LongArrow(220,130)(260,150)
  \Text(280,150)[]{$\ln q_+$}
  \LongArrow(120,130)(80,150)
  \Text(60,150)[]{$\ln q_-$}
  \Text(170,-30)[]{\large (b)}
\end{picture}}}
\end{minipage}
\caption{(a): A parton-parton scattering chain. Virtual
  links are denoted $k_i$ and real emissions $q_i$. In BFKL dynamics the
  transverse momenta are not ordered, and the result should be the same in any
  Lorentz frame. (b): The same chain in a ($y$, $\ln q_\perp^2$)
  plot. Final state radiation is allowed below the horizontal lines, and on the
  folds representing transverse jets.}
\label{fig:defqk}
\end{figure}

A reformulation of the CCFM model, called the Linked Dipole Chain model,
was presented in Ref.~\cite{Andersson:1995ju}. Here it was demonstrated 
that the inclusive cross section is fully
determined by a subset of the gluons in the CCFM approach, denoted
``$k_\perp$-changing'' gluons. In Fig.~\ref{fig:defqk}a,
we denote the real emitted gluons in a ladder $q_{\perp i}$, and
the virtual links $k_{\perp i}$. The $k_\perp$-changing emissions $k_{\perp i}$
are either much larger or much smaller than $k_{\perp i-1}$. This also means that
$q_{\perp i} \approx \max (k_{\perp i}, k_{\perp i-1})$. 
The chain in Fig.~\ref{fig:defqk}a is shown in the triangular phase space
diagram in Fig.~\ref{fig:defqk}b. The real gluons $q_i$ are ordered in $p_+$
and in $p_-$, and thus also in rapidity or angle. It was also demonstrated
that to get the full final states,
softer emissions have to be added below the horizontal lines in
Fig.~\ref{fig:defqk}b, as final state radiation. This also includes the folds
sticking out of the plane, which represent the transverse jets formed by the 
gluons $q_i$.

Thus in order to generate exclusive final states we should go through the
following steps:
\vspace{2mm}

1. Generate cascades for projectile and target

2. Determine which dipoles interact

3. Absorb non-interacting chains

4. Determine final state radiation

5. Hadronize
\vspace{2mm}

The main problems in this process are due to the large number of small dipoles
in the cascades. These have low cross sections, and are therefore not a 
big problem for inclusive cross sections. Because small dipoles correspond to
high transverse momenta, they do, however, have a large effect on the
properties of the final states. This implies that the result is sensitive to
details in the treatment of non-interacting dipoles.
Our aim is here therefore not to give very precise predictions, but rather 
to get insight into the dynamical features of small $x$ evolution and
saturation.

\begin{figure}
  \includegraphics[width = 0.5\linewidth]{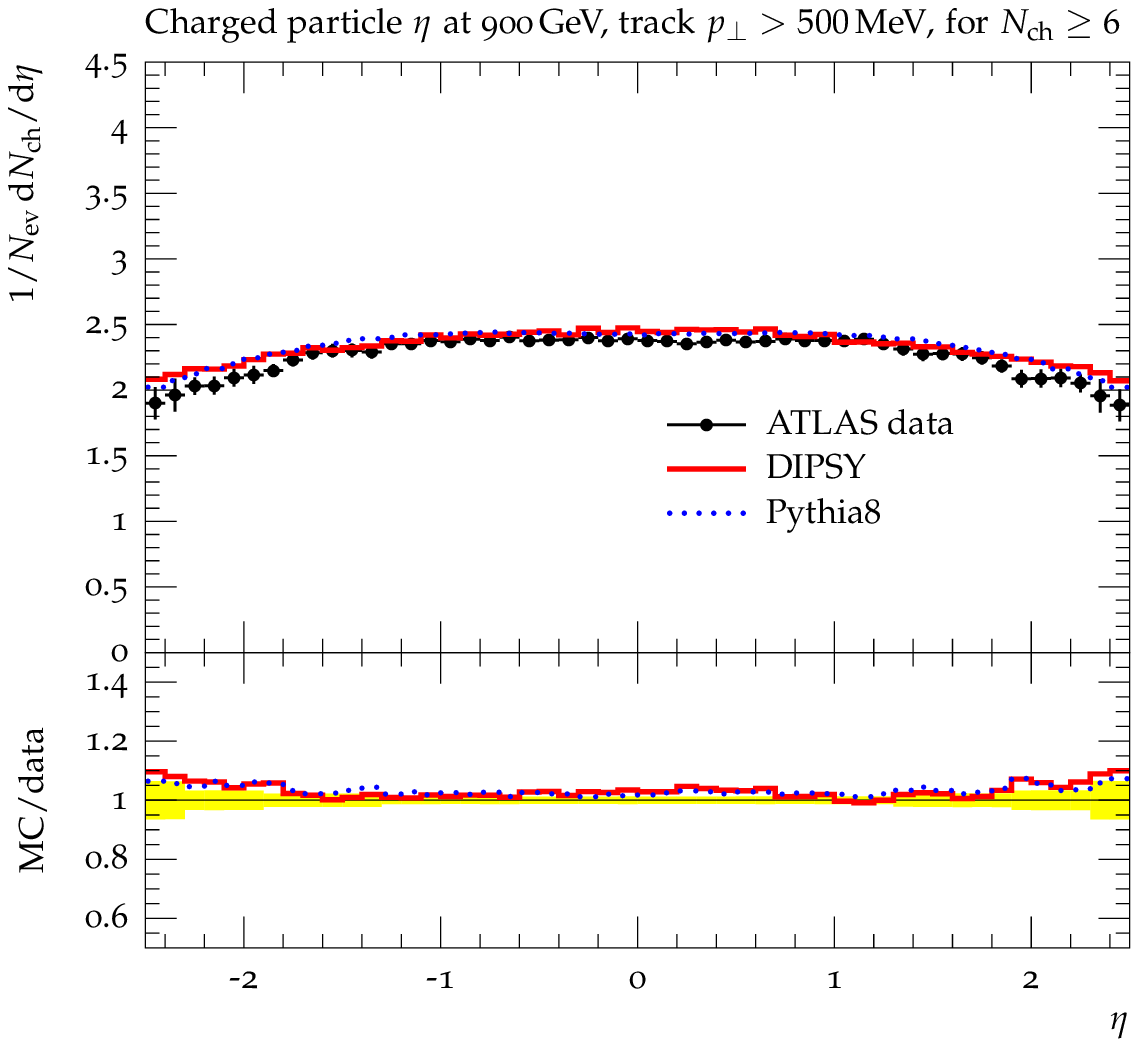}
\includegraphics[width = 0.5\linewidth]{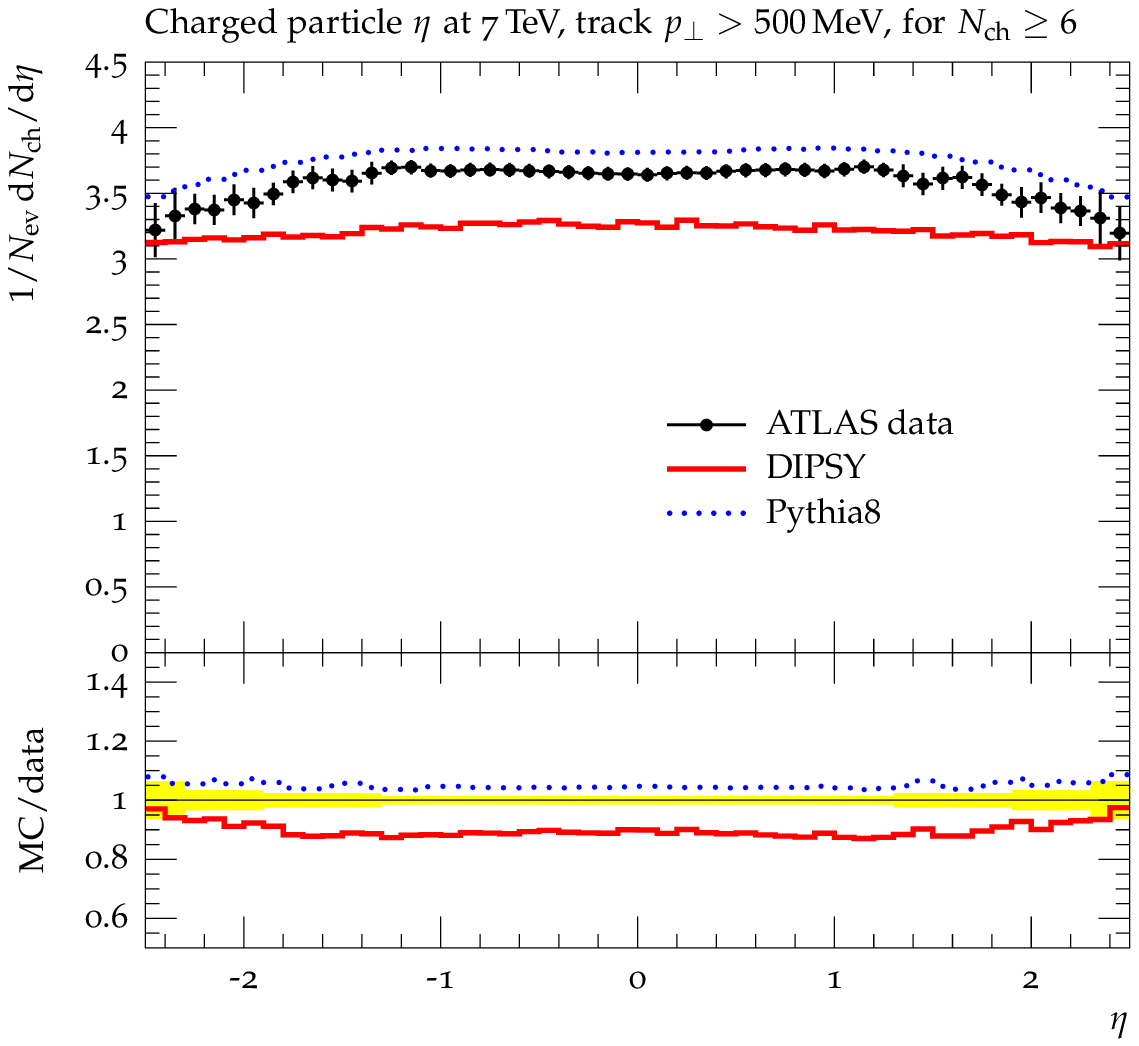}
\caption{$\eta$-distribution of charged particles at 0.9 and 7~TeV. 
The solid line shows the result from the DIPSY MC, and the dotted line 
\textsc{Pythia}. Data from the \textsc{Atlas} collaboration \cite{Aad:2010ir}.}
\label{fig:final1}
\end{figure}

\begin{figure}
\includegraphics[width = 0.5\linewidth]{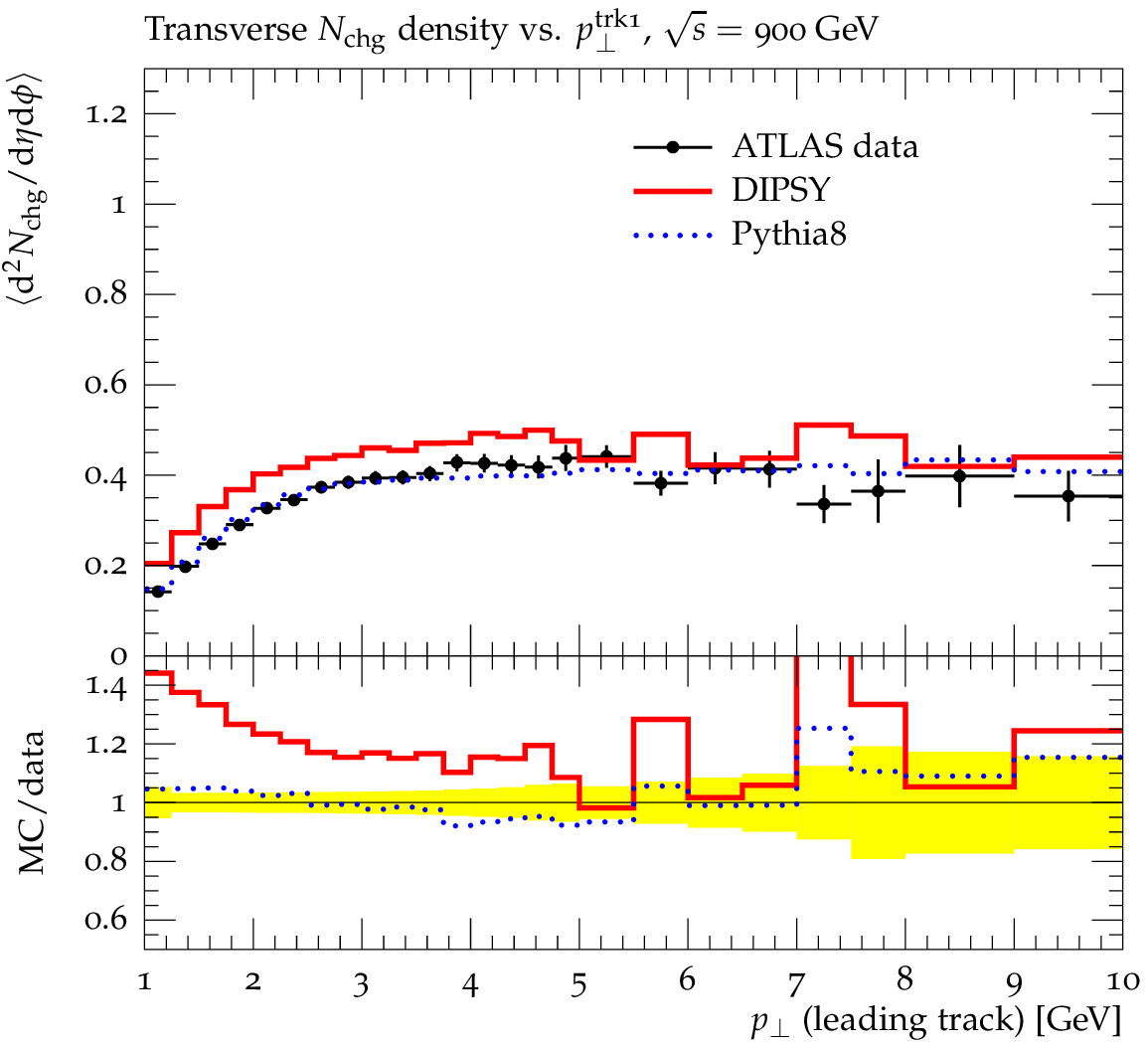}
\includegraphics[width = 0.5\linewidth]{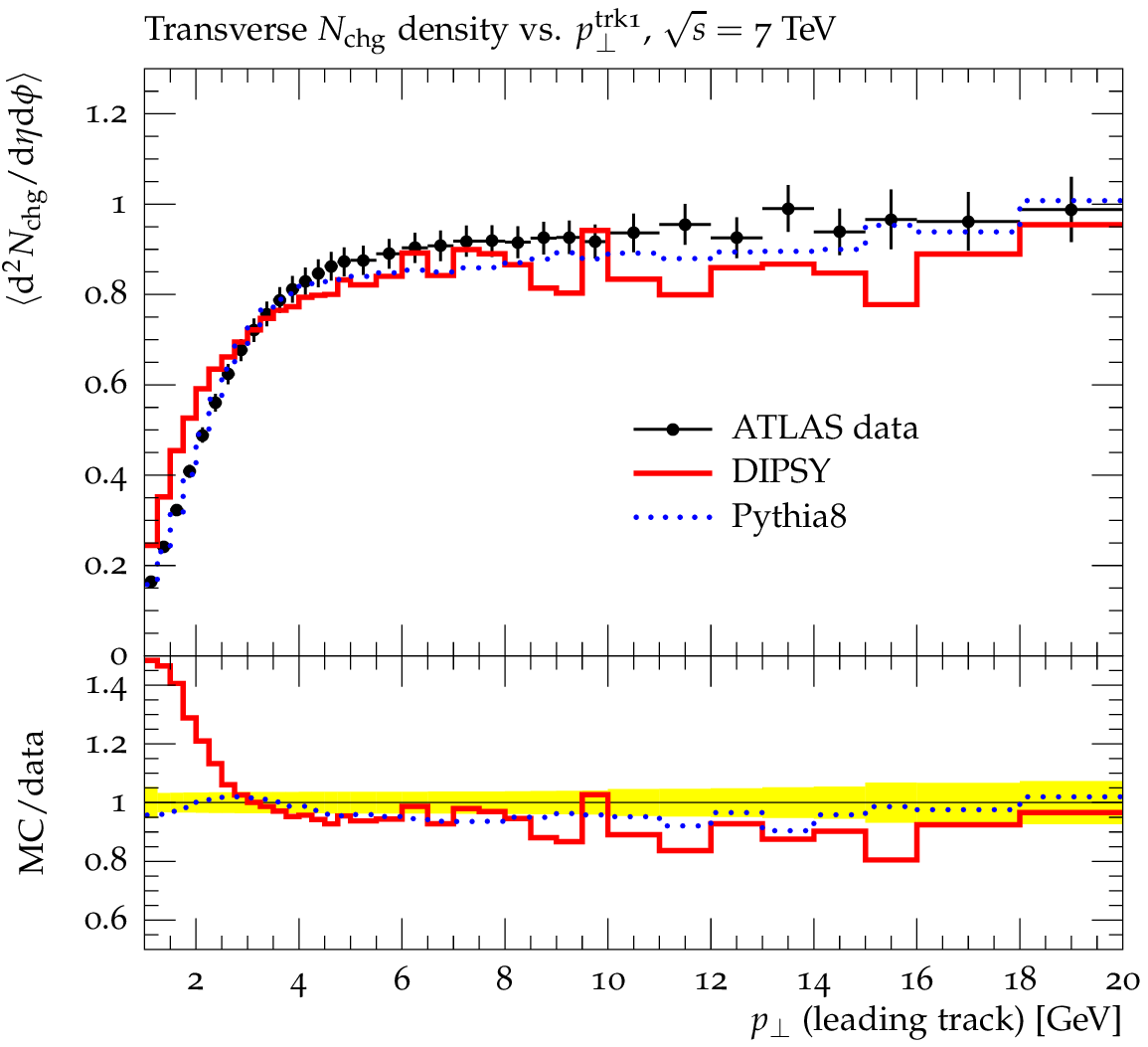}
\caption{$N_{ch}$ in transverse region \textit{vs} $p_\perp$ of leading
  charged particle. Notation as in Fig.~\ref{fig:final1}.
Data from the \textsc{Atlas} collaboration \cite{Aad:2010fh}.}
\label{fig:final2}
\end{figure}

As a few examples Figs.~\ref{fig:final1} and \ref{fig:final2} show comparisons
with ATLAS data for minimum bias and underlying events at 0.9 and 7 TeV. 
Fig.~\ref{fig:final1} shows the $\eta$-distribution of charged particles 
in minimum bias events. 
The solid line shows the result from the DIPSY MC, and the dotted line the 
result from \textsc{Pythia}. We note that the particle density
is well reproduced at 0.9 TeV, but does not grow fast enough with energy,
which is a problem also for other MCs which are not tuned individually for
each energy. The properties of the underlying event is shown in 
Fig.~\ref{fig:final2}, which presents the charged multiplicity in the
``transverse region'', as defined by Rick Field, as a function of the
$p_\perp$ of a leading charged particle. The model reproduces quite well
the increased density for higher $p_\perp$. More comparisons are found in
Ref.~\cite{Flensburg:2011kk}. 

\subsubsection{Nucleus collisions}

The model can also be applied to reactions
involving nuclei, where \emph{e.g.} saturation effects can be studied
including a proper geometry. Some early results are presented in
Ref.~\cite{arXiv:1108.4862}.  

\section{Summary}

In these lectures I have discussed interaction cross sections and particle
production in $e^+e^-$-ann., DIS, and high energy hadronic collisions. It
includes hadronization, initial and final state radiation, small $x$ 
evolution and saturation. 

I have also presented the Lund Dipole Cascade model for high energy
collisions, which is based on BFKL evolution and saturation. It is an
extension of Mueller's model, also including

\begin{itemize}
\item important non-leading effects in BFKL

\item saturation within the evolution

\item confinement

\item A MC implementation DIPSY

\end{itemize}

The model 
gives a good description of inclusive $pp$ and $ep$ cross sections 
(including diffraction),
and a fair description of exclusive final states (min. bias and underlying
event). It has fewer tunable parameters than other event generators,
and our aim is not to give very precise predictions, but rather 
to get insight into the dynamical features of small $x$ evolution and
saturation. As examples it is here possible to study effects of correlations, 
fluctuations, and finite transverse size in a way, which is not easy in other
approaches.

\end{document}